\documentclass{aa}

   \usepackage{graphicx}
   \usepackage{txfonts}

\begin{document}

\title{EVN observations of eleven GHz-Peaked-Spectrum radio sources at
       2.3/8.4GHz}
   \author{L. Xiang
          \inst{1}
          \and
           D. Dallacasa\inst{2,4}
           \and
           P. Cassaro\inst{3}
          \and
           D. Jiang\inst{5}
          \and
          C. Reynolds\inst{6}
          }

   \offprints{L. Xiang,
   \email{liux@ms.xjb.ac.cn}}

   \institute{National Astronomical Observatories, Chinese Academy of Sciences, 40-5 South Beijing Road, Urumqi 830011, China\\
             \email{liux@ms.xjb.ac.cn}
\and Dipartimento di Astronomia, Universita di Bologna, via Ranzani 1, I-40127 Bologna, Italy\\
             \email{dallacasa@ira.cnr.it}
         \and Istituto di Radioastronomia del CNR, C.P. 141, I-96017 Noto SR, Italy\\
             \email{cassaro@noto.ira.cnr.it}
         \and Istituto di Radioastronomia del CNR, via P. Gobetti 101, 40129 Bologna, Italy\\
\and Shanghai Astronomical Observatory, Chinese Academy of Sciences, 80 Nandan Road, Shanghai 200030, China\\
             \email{djiang@shao.ac.cn}
   \and Joint Institute for VLBI in Europe, Postbus 2, 7990 AA Dwingeloo, The Netherlands\\
             \email{reynolds@jive.nl}
             }

   \date{Received 3 August 2004 / Accepted 17 December 2004}

   \abstract{We present results of EVN observations of eleven
GHz-Peaked-Spectrum (GPS) radio sources at 2.3/8.4 GHz. These
sources are from the classical ''bright'' GPS source samples with
peak flux densities $>$ 0.2 Jy and spectral indices $\alpha <
-0.2$ ($S \propto \nu^{-\alpha}$) in the optically thick regime of
their convex spectra. Most of the target sources did not have VLBI
images at the time this project started. The aim of the work is to
find Compact Symmetric Object (CSO) candidates from the ''bright''
GPS samples. These CSOs play a key role in understanding the very
early stage of the evolution of individual radio galaxies. The
reason for investigating GPS source samples is that CSO candidates
are more frequently found among this class of radio sources. In
fact both classes, GPS and CSO, represent a small fraction of the
flux limited and flat-spectrum samples like PR+CJ1 (PR:
Pearson-Readhead survey, CJ1: the first Caltech--Jodrell Bank
survey) and CJF (the Caltech--Jodrell Bank flat spectrum source
survey) with a single digit percentage progressively decreasing
with decreasing flux density limit. Our results, with at least 3,
but possibly more CSO sources detected among a sample of 11,
underline the effectiveness of our approach. The three confirmed
CSO sources (1133+432, 1824+271, and 2121$-$014) are characterized
by a symmetric pair of resolved components, each with steep
spectral indices. Five further sources (0144+209, 0554$-$026,
0904+039, 0914+114 and 2322$-$040) can be considered likely CSO
candidates. The remaining three sources (0159+839, 0602+780 and
0802+212) are either of core-jet type or dominated by a single
component at both frequencies.

\keywords{galaxies: nuclei -- quasars: general -- radio continuum:
galaxies}
}

\titlerunning{EVN observations of eleven GHz-Peaked-Spectrum (GPS)
   radio sources at 2.3/8.4 GHz}
\authorrunning{L. Xiang et al.}
\maketitle

\section{Introduction}

Compact Symmetric Objects (CSOs) are a class of radio sources with
distinctive radio properties.
They are powerful and compact sources with overall size $<$ 1 kpc,
dominated by lobe/jet emission on both sides of the central engine,
and are thought to be relatively free of beaming effects (Wilkinson et
al. 1994). Their small size is most likely due to their youth ($<
10^4$ years) and not due to a dense confining medium (Owsianik \&
Conway 1998). A unification scenario assumes that CSOs evolve into
Medium-size Symmetric Objects (MSOs, 1-15 kpc), which, in turn,
evolve into Large Symmetric Objects (LSOs, $>$ 15 kpc), i.e. large
FRII radio sources (Fanti et al. 1995, Readhead et al. 1996, Taylor
2003).

CSOs are of particular interest in the study of the physics and
evolution of active galaxies. First, CSO activity may be triggered
through galaxy-galaxy merging, and the host galaxies of some CSOs
exhibit merging or interaction diagnostics (Xiang et al. 2000,
Perlman et al. 2001). A few CSOs have been found in known merging
systems, e.g. 1345+125 (Xiang et al. 2002, Lister et al. 2003).
Second, hot-spot proper motions from 0.1$c$ to 0.8$c$ have been
detected in CSOs (Fanti 2000 and references therein, Polatidis \&
Conway 2003, Kellermann et al. 2004), and the estimated ages are
between 100 and a few thousand years. A statistical study of the
hot-spot proper motions and ages will be useful to pursue the CSO
evolution from the very beginning of its radio burst. Third, CSOs
inhabit the densest parts of their host galaxies (in other words,
the nuclear regions of galaxies), and as such provide excellent
probes of the ambient medium of the active galactic nuclei through
e.g. HI absorption, free-free absorption (FFA) and the Thomson
scattering effect (Taylor 2003, Pihlstr\"om et al. 2003, Xiang
2004).

However, only a few CSOs have been well explored, which is not
enough for a statistically sound study. The first systematic
search for CSOs among the population of VLBI radio sources was
carried out on the Pearson and Readhead (PR, Pearson \& Readhead
1988) and the first Caltech--Jodrell Bank (CJ1, Xu et al. 1995)
VLBI surveys. Together they form a flux limited complete sample of
200 sources with $S_{5GHz}>$ 0.7 Jy and $\delta>35^{\circ}$. A
total of 14 CSOs were found (Polatidis et al. 1999) with a
fractional incidence of 7\%. If we consider also the second
Caltech--Jodrell Bank VLBI survey (CJ2, Henstock et al. 1995) and
the Caltech--Jodrell Bank Flat spectrum source survey (CJF, Taylor
et al. 1996), the fraction goes down to 4.4\% (18 CSOs out of 411
radio sources). If we finally consider also the VLBA Calibrator
Survey we obtain 39 CSOs out of 1900 radio sources (2.1\%).

The CSO detection rate appears to decrease as the sample size
increases by lowering the flux density limit, i.e. there seems to
be a trend to lower CSO incidence among fainter sources. On the
other hand, the CJ2, CJF and VCS samples are defined as flat
spectrum sources with spectral indices $\alpha <$ 0.5 (we use $S
\propto \nu^{-\alpha}$ in this paper). In contrast, a half of the
CSOs in the flux limited complete sample, PR+CJ1, show steep
spectra of $\alpha >$ 0.5. Therefore, steep spectrum CSOs will be
missed in the flat spectrum samples.

GPS (GHz-Peaked-Spectrum) radio sources are a significant fraction
of the bright (centimeter-wavelength-selected) radio source sample
($\approx$ 10\%) but they are not well understood. The GPS sources
are powerful ($P_{1.4GHz}\geq10^{25}WHz^{-1}$), compact ($\leq$ 1
kpc) and have convex radio spectra that peak between about 0.5 and
10 GHz (observer's frame). Only about 12\% of GPS sources show
extended radio emission ($>$ 1 kpc), and it is diffuse and very
faint. Most GPS sources appear to be truly compact and isolated
(see O'Dea 1998 for a review of the GPS source properties).

It is found that, of 14 detected CSOs in the PR+CJ1 sample, 10 are
GPS radio sources. Furthermore, six out of 10 GPS-type CSOs have
shown steep spectra with spectral indices of $\alpha
>$ 0.5 in their optically thin regime, and their cores are
relatively weak. Given that these CSOs are small radio sources, it
is quite likely that their low frequency radio emission will be
absorbed due to either Synchrotron Self Absorption (SSA), or
Free-Free absorption, giving rise to a peaked (GPS) radio
spectrum. This is why a number of CSOs in the PR+CJ1 sample are
GPS sources. Hence, searching in GPS samples would be an efficient
way to find CSO sources.

GPS radio sources are usually identified with galaxies at low to
intermediate redshift ($z < 1$), while quasars are at higher
redshift (Snellen et al. 1999). Some GPS radio sources have no
optical counterparts yet, and these empty fields are likely to be
distant galaxies, too faint to be detected. A strong correlation
between the milliarcsecond morphology and the optical host in
bright GPS sources suggests that galaxies are generally associated
with CSOs while quasars have more often a core-jet or complex
morphology (Stanghellini et al. 1998).

There are a few tens of sources in bright GPS source samples, and
a few of them have never (or poorly) been imaged with VLBI. From
the lists of the bright GPS source samples (O'Dea et al. 1991,
1996, de Vries et al. 1997), we have selected a number of sources
to be imaged with VLBI. The first observation run was made at 1.6,
2.3 and 8.4 GHz with the EVN and MERLIN for 10 sources. From these
multi-frequency high resolution images we found 2 CSOs and 4 CSO
candidates (Xiang et al. 2002).

In the following we present results from the second run with the
EVN at 2.3 GHz and at 8.4 GHz. We observed the sources with peak
flux densities $>$ 0.2 Jy, spectral indices $\alpha < -0.2$ in the
optically thick regime of their convex spectra, and  absent or
insufficient morphological information in the literature, so that
a proper classification was not available.

Our search for CSOs from GPS samples is a complementary to that of
the COINS group (Peck \& Taylor 2000) which looks for CSOs based
on samples consisting mainly of flat spectrum radio sources.

\section{Observations and data reduction}

The observations were carried out on 4 June 2002 at 2.3/8.4 GHz
using the MKIV recording system with a bandwidth of 16 MHz at each
frequency, in right circular polarization. The EVN antennae
participating in this experiment were those located at Effelsberg,
Wettzel, Medicina, Noto, Matera, Onsala, Yebes, Urumqi and
Shanghai. All stations produced useful fringes. Snapshot
observations of the 11 target sources (see Table 1) for a total of
24 hours were carried out. OQ208 and BL Lac were observed as
calibration sources. In January 2003, the data correlation was
performed at the VLBI correlator at the Joint Institute for VLBI
in Europe in Dwingeloo.

The Astronomical Image Processing System (AIPS) software was used
for editing, a-priori calibration,  fringe-fitting,
self-calibration and imaging. The AIPS procedure UVCRS was used to
derive the antenna gains of Wettzel and Matera for which
appropriate system temperature measurements were not available.

The absolute flux density scale is estimated to be accurate to
about 10\% based on the calibrator source OQ208. The amplitude of
OQ208 on the shortest baseline Effelsberg-Wettzel was found to be
1.75 Jy at 2.3 GHz, and 2.0 Jy at 8.4 GHz in our calibrated data.
The source is not resolved on this baseline, and these values are
consistent with 1.65 Jy at 2.3 GHz, and 1.95 Jy at 8.4 GHz
(Stanghellini et al.  1998) within an error of 5\%. This source is
believed not to have varied substantially at 2.3 GHz and at 8.4
GHz in recent years (Stanghellini et al. 1997); the flux density
at 8.55 GHz measured at Effelsberg in May 2001 is 2.03$\pm{0.10}$
Jy (A. Kraus, priv. communication). For most sources we obtained
the final images using near-natural weighting of the visibilities
in order to have a better sensitivity to larger source components.

\section{Results and comments on sources}

In this section we present and briefly highlight the results from
the dual frequency VLBI observations.

Some basic information on the target sources is presented in Table
1. The source component parameters as measured on the VLBI images
are summarized in Table 2. In the following we consider a possible
interpretation of each source based on the morphology and the
spectral properties derived from the data presented here.

   \begin{table*}
   \label{first}
   \caption[]{GPS sources. Columns 1 through 14 provide source name,
   optical identification (G: galaxy, Q: quasar, EF: empty field; a
   '?' means that the classification is uncertain), optical magnitude
   and filter, redshift (* is a photometric estimate by Heckman
   et al. 1994), linear scale factor pc/mas
         [$H_{0}$=100km/s and $q_{0}=0.5$ have been assumed],
   maximum VLBI angular size in mas, maximum VLBI linear size in pc,
   2.7 GHz flux density in Jy from the NASA/IPAC Extragalactic Database
   (NED), source spectral index from our VLBI images, low frequency
   (optically thick) spectral index, high frequency (optically thin)
   spectral index, turnover frequency in GHz, peak flux density in Jy
   and references for the spectral information (ref: 1, de Vries et
   al. 1997; 2, Stanghellini et al. 1998).}
         $$
         \begin{array}{p{0.1\linewidth}*{5}{p{0.06\linewidth}}p{0.06\linewidth}*{8}{p{0.06\linewidth}}}
            \hline
            \hline
            \noalign{\smallskip}
source & id & m & z & pc/mas & $\theta$ & L & $S_{2.7}$ & $\alpha_{vlbi}$ & $\alpha_{l}$ & $\alpha_{h}$ & $\nu_{m}$ & $S_{m}$ & ref \\
& & & & & mas & pc & Jy & & & & GHz & Jy & \\
            \noalign{\smallskip}
            \hline
            \noalign{\smallskip}

\object{0144+209} & EF  &        &       &      & 50   &      & 0.85 & 0.87 & -0.63  & 0.59  & 1.3 & 1.2 & 1 \\
\object{0159+839} & Q?  & 17V    &       &      & 90   &      &      & -0.38 & -0.47  & 0.91  & 5.0 & 0.2 & 1 \\
\object{0554$-$026} & G   & 18.5V  & 0.235 & 2.37 & 30   & 71   & 0.62 & 1.25 & -1.07  & 0.63  & 1.0 & 0.8 & 1 \\
\object{0602+780} & G   & 22.0R  & 1.1*  & 4.3  &$<$25 &$<$108 &      & -0.57 & -0.58  & 0.57  & 5.0 & 0.2 & 1 \\
\object{0802+212} & G   & 22.5R  &       &      & 25   &      & 1.03 & 0.50 & -0.54  & 0.57  & 1.8 & 1.0 & 1 \\
\object{0904+039} & G   & 22.1I  &       &      & 80   &      & 0.44 & 1.72 & -0.23  & 0.83  & 0.8 & 1.0 & 1 \\
\object{0914+114} & G   & 20.0r  & 0.178 & 1.95 & 150  & 293  & 0.31 & 2.0 & -0.1   & 1.6   & 0.3 & 2.3 & 2 \\
\object{1133+432} & EF  &        &       &      & 40   &      &      & 1.37 & -0.60  & 0.6   & 1.0 & 1.4 & 1 \\
\object{1824+271} & G?  & 22.9R  &       &      & 45   &      & 0.23 & 1.43 & -0.39  & 0.75  & 1.0 & 0.4 & 1 \\
\object{2121$-$014} & G   & 23.3R & 1.158 & 4.3  & 88   & 378   & 0.64 & 1.11 & -0.56  & 0.75  & 0.5 & 1.8 & 1 \\
\object{2322$-$040} & G   & 23.5R &       &      & 65   &       & 0.91 & 1.36 & -0.42  & 0.75  & 1.4 & 1.3 & 1 \\

  \noalign{\smallskip}
  \hline
  \end{array}
         $$
  \end{table*}

  \begin{table*}
  \label{second}
  \caption[]{Parameters of the source components in the VLBI images at
  2.3 and 8.4 GHz. The various columns provide: (1)
         source name and possible classification (CSOc: Compact Symmetric Object candidate, cj: core-jet), (2)
         total image intensity measured with IMEAN at 2.3 GHz, (3) component identification, (4) spectral index
         of component computed with \textit{Sint}, (5),(6) peak (\textit{Sp}) and
         integral intensity (\textit{Sint}) of component at 2.3 GHz measured with JMFIT, (7),(8) major/minor axes and position angle
         of component at 2.3 GHz measured with JMFIT, (9) total image intensity measured with IMEAN at 8.4 GHz, (10) component
         identification, (11),(12) peak and integral intensity of component at 8.4 GHz measured with JMFIT, (13),(14)
         major/minor axes and position angle of component at 8.4 GHz measured with JMFIT.}
         $$
         \begin{array}{p{0.1\linewidth}*{5}{p{0.05\linewidth}}*{1}{p{0.1\linewidth}}*{5}{p{0.05\linewidth}}*{1}{p{0.1\linewidth}}*{1}{p{0.05\linewidth}}}
          \hline
           \hline
            \noalign{\smallskip}
Name & $S_{2.3}$& & $\alpha_{c}$ & \textit{Sp} & \textit{Sint} & $\theta_{1}\times\theta_{2}$ & PA         & $S_{8.4}$ & & \textit{Sp} & \textit{Sint} & $\theta_{1}\times\theta_{2}$ & PA \\
class& mJy      & &              &mJy & mJy  & mas          & $^{\circ}$ & mJy       & & mJy& mJy  & mas         & $^{\circ}$\\
            \noalign{\smallskip}
            \hline
            \noalign{\smallskip}

\object{0144+209} & 903   &  &     &      &      &  &                  &293& A & 47 & 67.8 & 2.2$\times$0.7 & 132  \\
    CSOc &       &  &     &      &      &  &                  &   & B & 113  & 168 & 2.3$\times$0.8 & 155 \\
         &       &  &     &      &      &  &                  &   & C & 39  & 80 & 4.0$\times$0.8 & 137 \\

\object{0159+839} &  23   &A &-0.86& 12   & 12.2 & 4.6$\times$4.6 & 0  & 38& A & 31 & 37 & 1.1$\times$0.9 & 106\\
 cj      &       &B &     & 9.0  & 21.6 & 19$\times$3.4  & 168&   &   &     &       &   &\\
         &       &C &     &4.3   & 5.8  & 7.4$\times$3.4 & 148&   &   &     &       &   &\\

\object{0554$-$026} & 275   &A &0.69 &80.2  & 116.6& 5.6$\times$2.2 & 6.3& 55& A1 & 15& 35& 2.3$\times$0.9 & 29\\
 CSOc    &       &  &     &      &      &                &    &   & A2 & 7.7& 13 & 3.5$\times$0.5  & 6\\
         &       &B & 1.19& 76.2 & 129.5& 7.1$\times$2.6 & 13 &   & B  & 11 & 28& 1.5$\times$1.1 & 63\\

\object{0602+780} & 50    &  &     &44.1  & 48.6 & 4.8$\times$3.0 & 112&105&   & 93& 104 & 0.8$\times$0.6 & 127 \\
 cj      &&&&&&&&&&&&&\\

\object{0802+212} & 755   &  &   &      &      &           &           & 409& A & 263& 293& 0.5$\times$0.3& 7.8\\
      cj &       &  &   &      &      &          &            &    & B & 71 & 140& 1$\times$1 & 144\\

\object{0904+039} & 508 &A &    & 271 & 367  & 11$\times$2.7      & 37 & 55 & A1& 17 & 24& 1.6$\times$1.1 & 138\\
 CSOc    &     &B &    & 82  & 112  & 11$\times$5.2      & 22 &    & A2& 7.4 & 15 & 2.6$\times$1.3& 95\\
         &     &C &    &23 & 56 & 32$\times$9.7          & 113&    & A3&  3.1& 8.6 & 3.5$\times$2.0 &99 \\

\object{0914+114} & 321 &A &    &43 &46  & 5$\times$5         & 0    &      &   &    &    &            & \\
 CSOc    &     &B &    &22 & 31 & 6.5$\times$5.3     & 102  &      &   &    &    &            &\\
         &     &C &1.73&190& 214& 4$\times$1.8       & 96   &  24  & C?& 20 & 23 &0.9$\times$0.7 & 158\\
         &     &D &    &18.5 & 27 & 7.7$\times$3.7   & 99   &      &   &    &    &              &\\

\object{1133+432} & 1068&A & 1.23&557& 692 & 3.4$\times$1.5   & 92   &  183 & A & 116   & 141  &1.2$\times$0.9 & 74\\
 CSO     &     &B &1.65 &282  & 367 & 3.8$\times$2.2 & 108  &      & B & 28 & 44 & 1.9$\times$1.7 & 128\\

\object{1824+271} & 222  &A & 1.37&118& 163& 6.4$\times$1.2   & 97    &  35 & A & 24& 28 & 1.7$\times$1.1 & 116\\
 CSO     &      &B &1.63 &31 & 58   &12$\times$12    & 0     &     & B & 5.4  & 7.1  & 2.0$\times$1.4  & 129\\

\object{2121$-$014} & 604 &A &1.43 &236& 253 & 2.9$\times$0.6   & 113   & 145 & A & 26 & 40 & 3.5$\times$2.6& 81\\
 CSO     &     &B &     &43 & 164   &  27$\times$1.3 & 108   &     &   &    &       &    &\\
         &     &C &0.65 &151& 237 & 7.8$\times$4.1   & 101   &     & C & 67 & 103   &  3.6$\times$1.6  &46\\

\object{2322$-$040} & 956 &A &1.1 &536 & 662 & 3.2$\times$1.6   & 109   & 166 & A & 123 & 160 & 1.4$\times$0.5 & 102\\
 CSOc    &     &B &    &129 & 234 & 6.7$\times$2.2   & 109   &     & ?  &       &              &   & \\

           \noalign{\smallskip}
            \hline
         \end{array}
         $$
   \end{table*}

\subsection{\object{PKS 0144+209} (\object{JVAS 0146+2110}, \object{NVSS J014658+211024})}

This radio source does not have an optical counterpart and is
reported as an empty field in Table 1. Snellen et al. (2002)
estimated its APM (the Automated Plate Measurement Facility at
Cambridge) red magnitude is $>$ 20, and the APM blue magnitude is
$>$ 22 based on the digitized Palomar Observatory Sky Survey
(POSS) plates.

In the Jodrell Bank--VLA Astrometric Survey (JVAS), the total flux
density at 8.4 GHz for the source is 337.2 mJy and the source is
classified as suitable (i.e. pointlike at subarcsecond resolution)
as a MERLIN phase calibrator (Wilkinson et al. 1998). It was not
detected at 22 GHz with the single dish in Mets\"ahovi
(Ter\"asranta et al. 2001).

\begin{figure}
  \includegraphics[width=5.6cm]{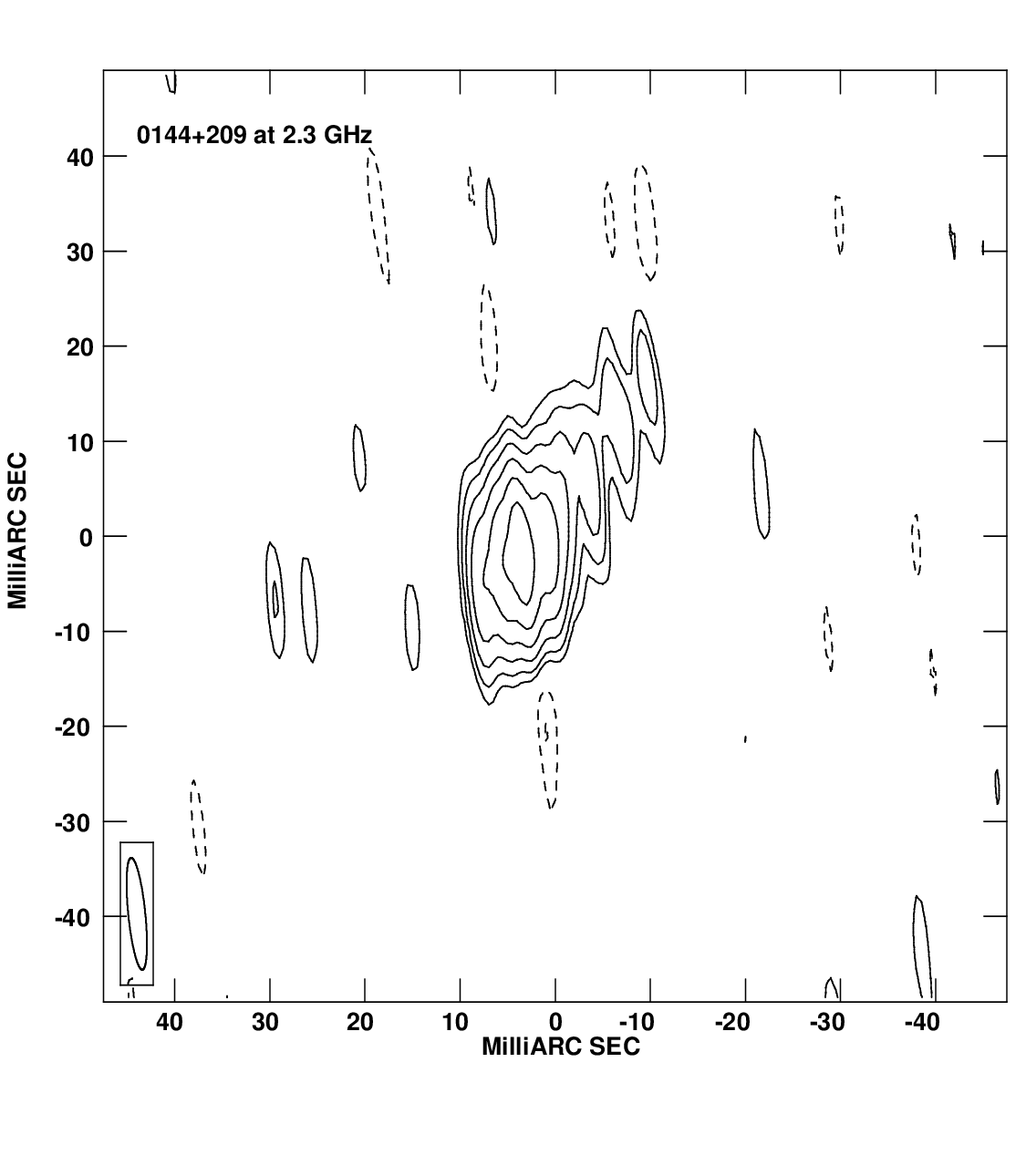}
  \caption{0144+209 at 2.3 GHz, the restoring beam is $11.8\times1.8$
      mas in PA $5.8^{\circ}$, the rms noise on the image is 2
      mJy/beam, the lowest contour is 2.5$\sigma$, the peak flux
      density is 286 mJy/beam. The contour levels here and the
      following figures are -1, 1, 2, 4, 8, 16, 32, 64, 128, 256, 512.}
      \label{0144S}
   \end{figure}

\begin{figure}
      \includegraphics[width=5.6cm]{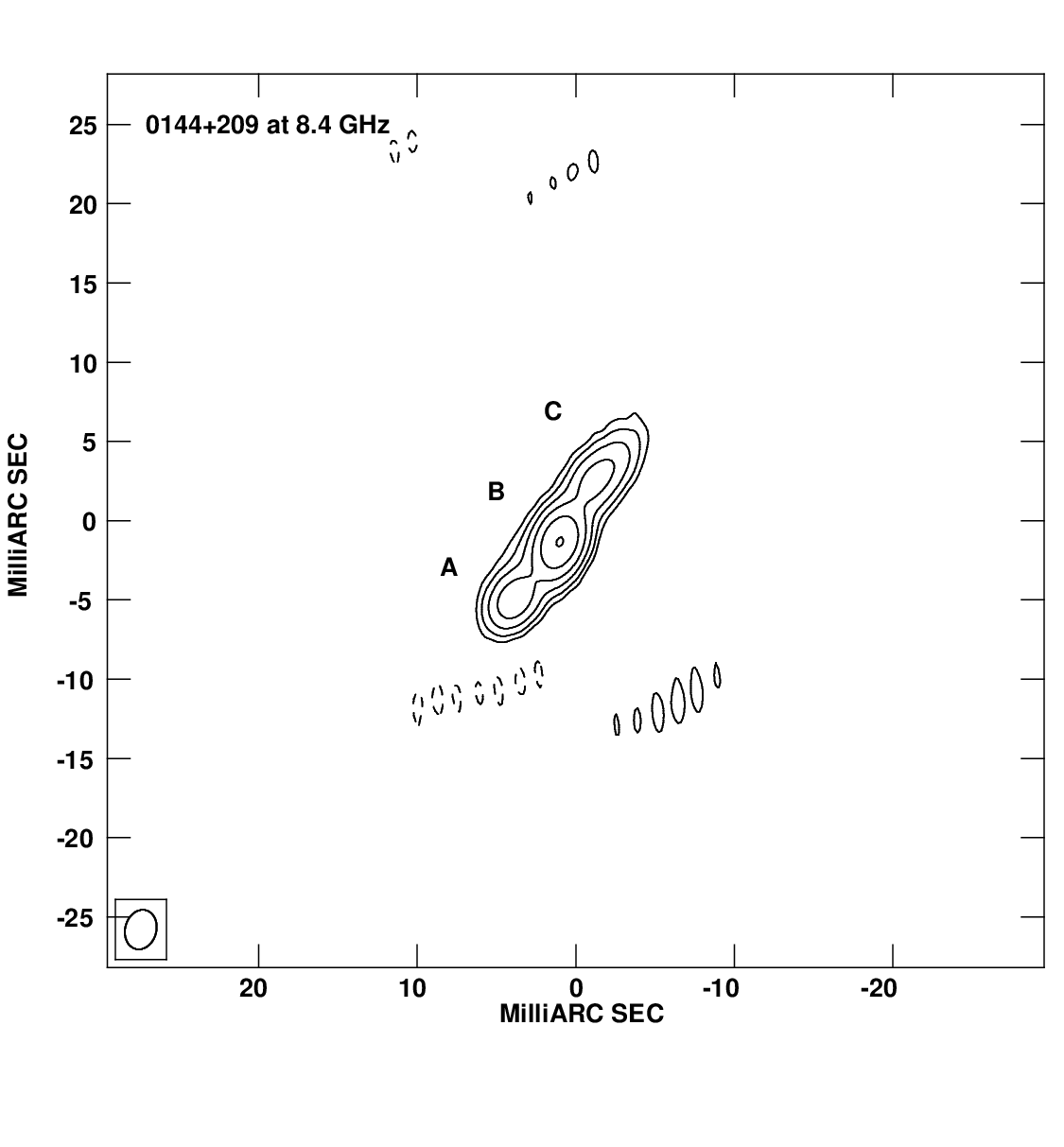}
      \caption{0144+209 at 8.4 GHz, the restoring beam is $2.5\times2.0$ mas in PA $-15.4^{\circ}$, the rms noise
      on the image is 1.0 mJy/beam, the lowest contour is 3.5$\sigma$, the peak flux density is 114 mJy/beam.}
      \label{0144X}
   \end{figure}

The VLBI images exhibit an elongated structure with a position
angle of $\sim -40^\circ$. At 2.3 GHz (Fig. \ref{0144S}), the
total source size detected is more extended to the North-West than
at 8.4 GHz (Fig. \ref{0144X}). The source is also found in the
VLBA Calibrator Survey (VCS, Beasley et al. 2002). A 2.3 GHz image
is available from the VCS web page, which is similar in structure
to our image. The source is a core-jet source if its southernend
is the core of the source. However, if the strongest central
component is the core, the source could be a CSO instead. The
total flux density accounted for in our image at 8.4 GHz is $\sim$
90\% of the JVAS measurement.

\subsection{\object{0159+839} (\object{JVAS J0207+8411}, \object{NVSS J020713+84111})}

Spoelstra et al. (1985) classified this object as a 17th magnitude
'stellar' object, de Vries et al. (2000) marked the object as a
'quasar?'. This is a ROSAT source, \object{1RXS J020716.5+841126}.

\begin{figure}
      \includegraphics[width=5.6cm]{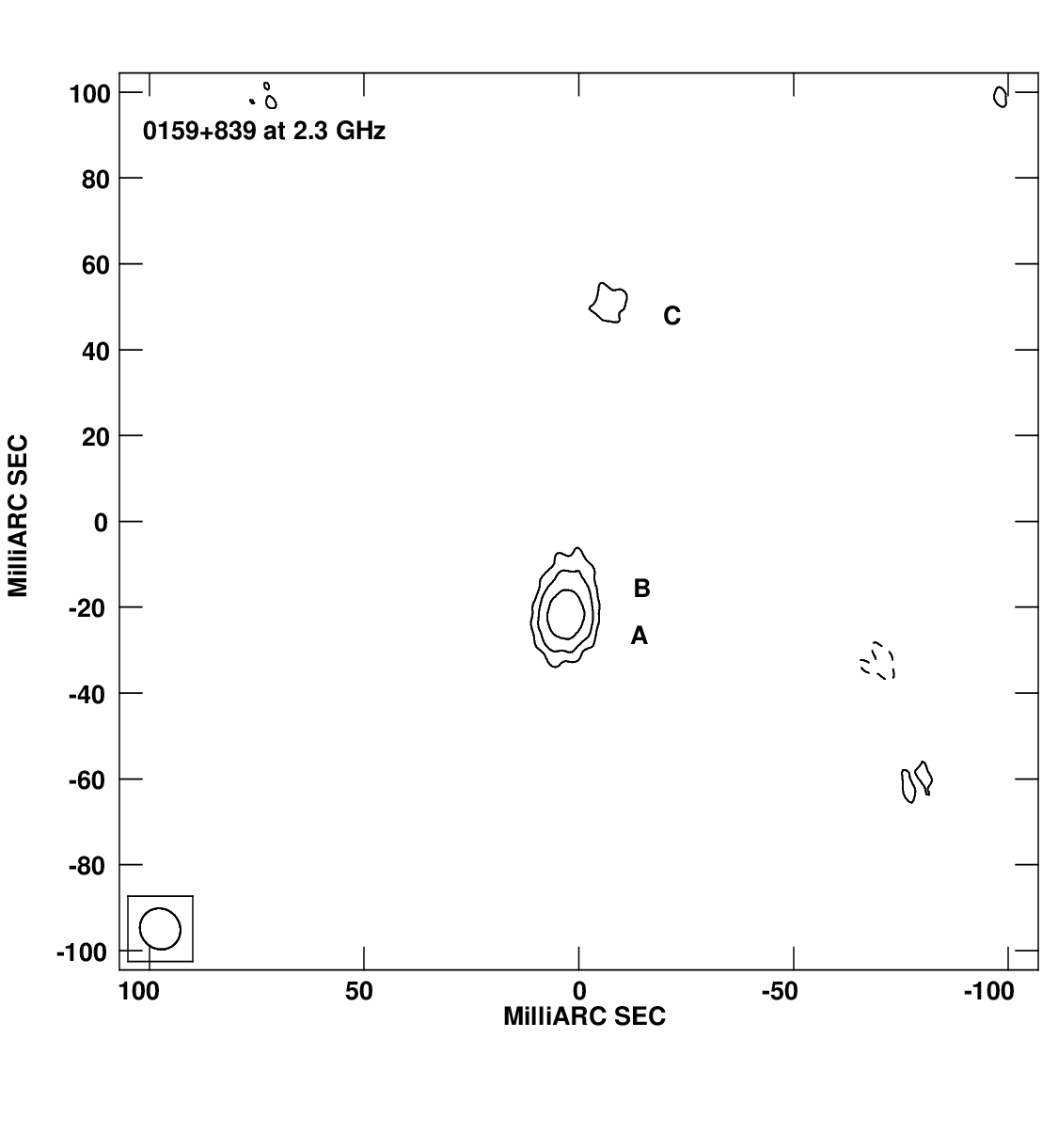}
      \caption{0159+839 at 2.3 GHz, the restoring beam is $9.8\times9.2$ mas in PA $39.3^{\circ}$, the rms noise
      on the image is 0.8 mJy/beam, the lowest contour is 3.7$\sigma$, the peak flux density is 20.5 mJy/beam.}
      \label{0159S}
   \end{figure}

\begin{figure}
      \includegraphics[width=5.7cm]{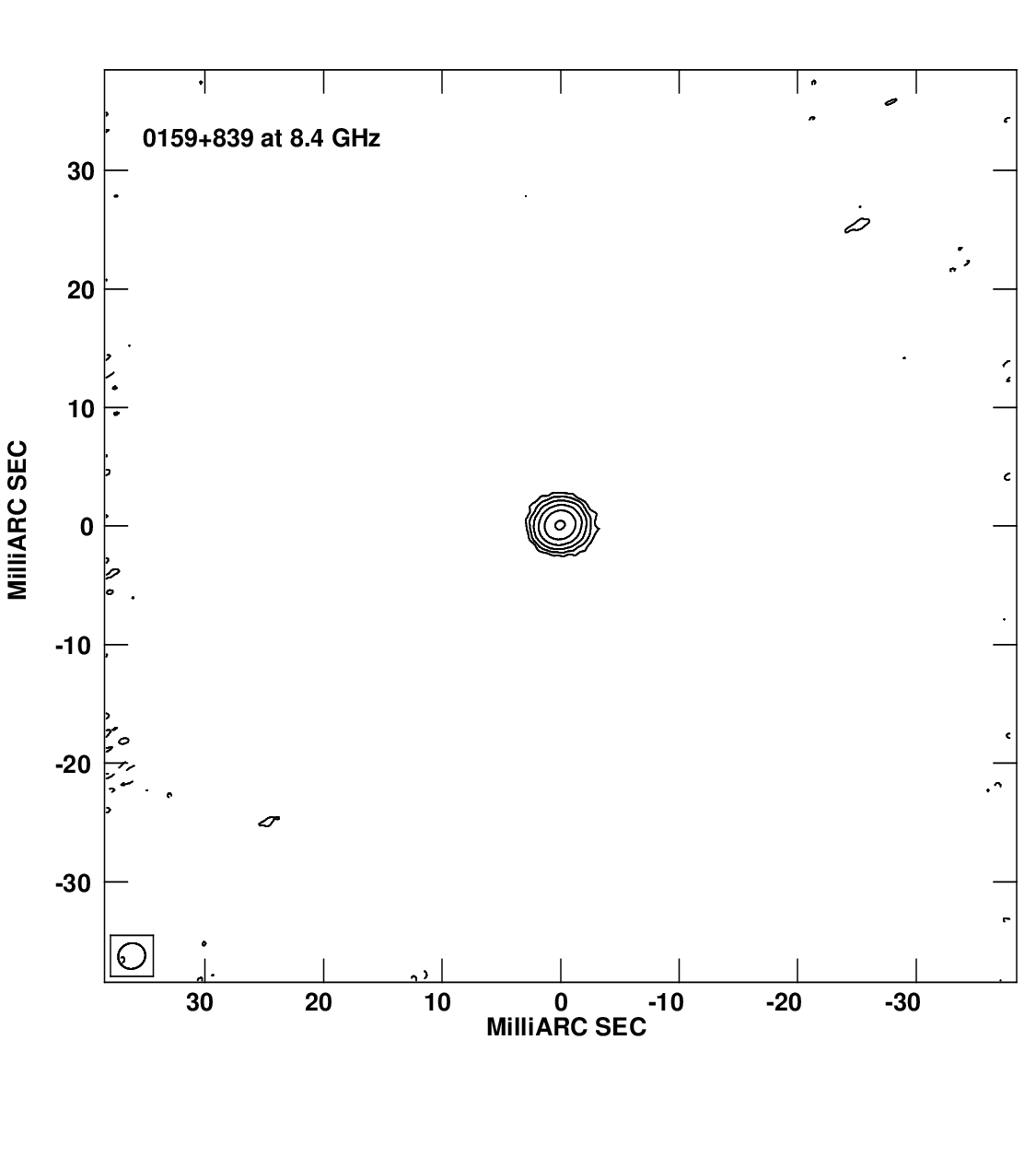}
      \caption{0159+839 at 8.4 GHz, the restoring beam is $2.3\times2.1$ mas in PA $-67.6^{\circ}$, the rms noise
      on the image is 0.3 mJy/beam, the lowest contour is 2.0$\sigma$, the peak flux density is 31 mJy/beam.}
      \label{0159X}
   \end{figure}

In the radio, the JVAS flux density at 8.4 GHz is 77.9 mJy and the
source is a suitable MERLIN phase calibrator (Wilkinson et al.
1998). Our VLBI images show a weak, slightly resolved source at
2.3 GHz (Fig.\ref{0159S}), and a point-like object at 8.4 GHz
(Fig. \ref{0159X}, where about a half of the JVAS flux density is
accounted for). The source is resolved into core-jet like
(components A,B), and a hint of a third component C in the
direction of the jet, while it is unresolved at 8.4 GHz. Further,
the VLBI structure as a whole has an inverted spectral index
($\alpha = -0.38$). The spectrum becomes even more inverted
($\alpha = -0.86$) if we consider that the component seen at 8.4
GHz corresponds to the more compact component (A) at 2.3 GHz. A
possible explanation for the different morphologies at 2.3 GHz and
at 8.4 GHz is that the components A and B represent a core-jet
source, with only the inverted spectrum core component being
detected at 8.4 GHz. The jet component, with a more optically thin
synchrotron spectrum, is undetected at 8.4 GHz due to the low
dynamic range of the image.

\subsection{\object{PKS 0554$-$026} (\object{NVSS J055652$-$024105})}

\begin{figure}
      \includegraphics[width=6cm]{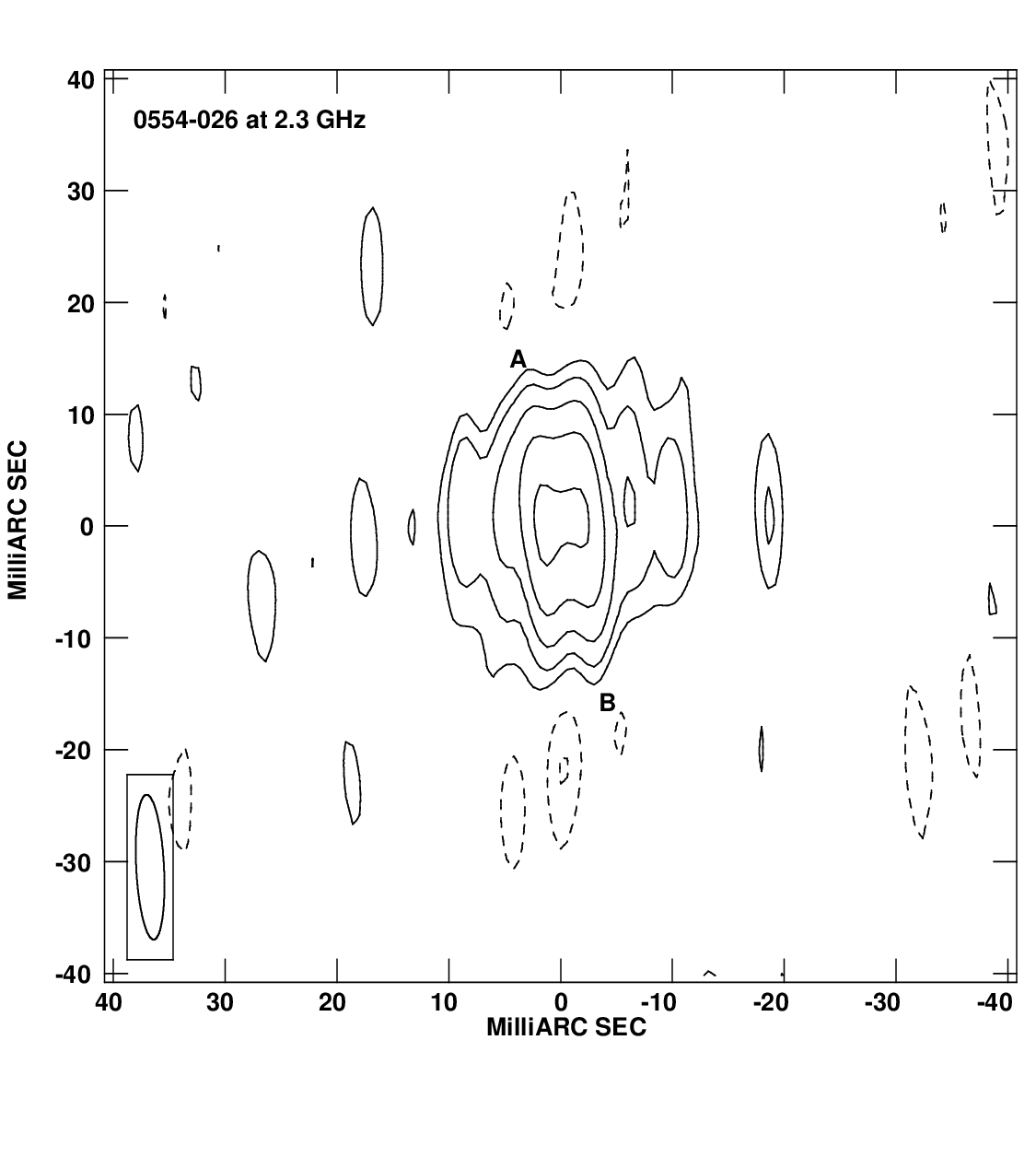}
      \caption{0554$-$026 at 2.3 GHz, the restoring beam is $12.8\times2.5$ mas in PA $2.9^{\circ}$, the rms noise
      on the image is 2.2 mJy/beam, the lowest contour is 2$\sigma$, the peak flux density is 86 mJy/beam.}
      \label{0554S}
   \end{figure}

\begin{figure}
      \includegraphics[width=6cm]{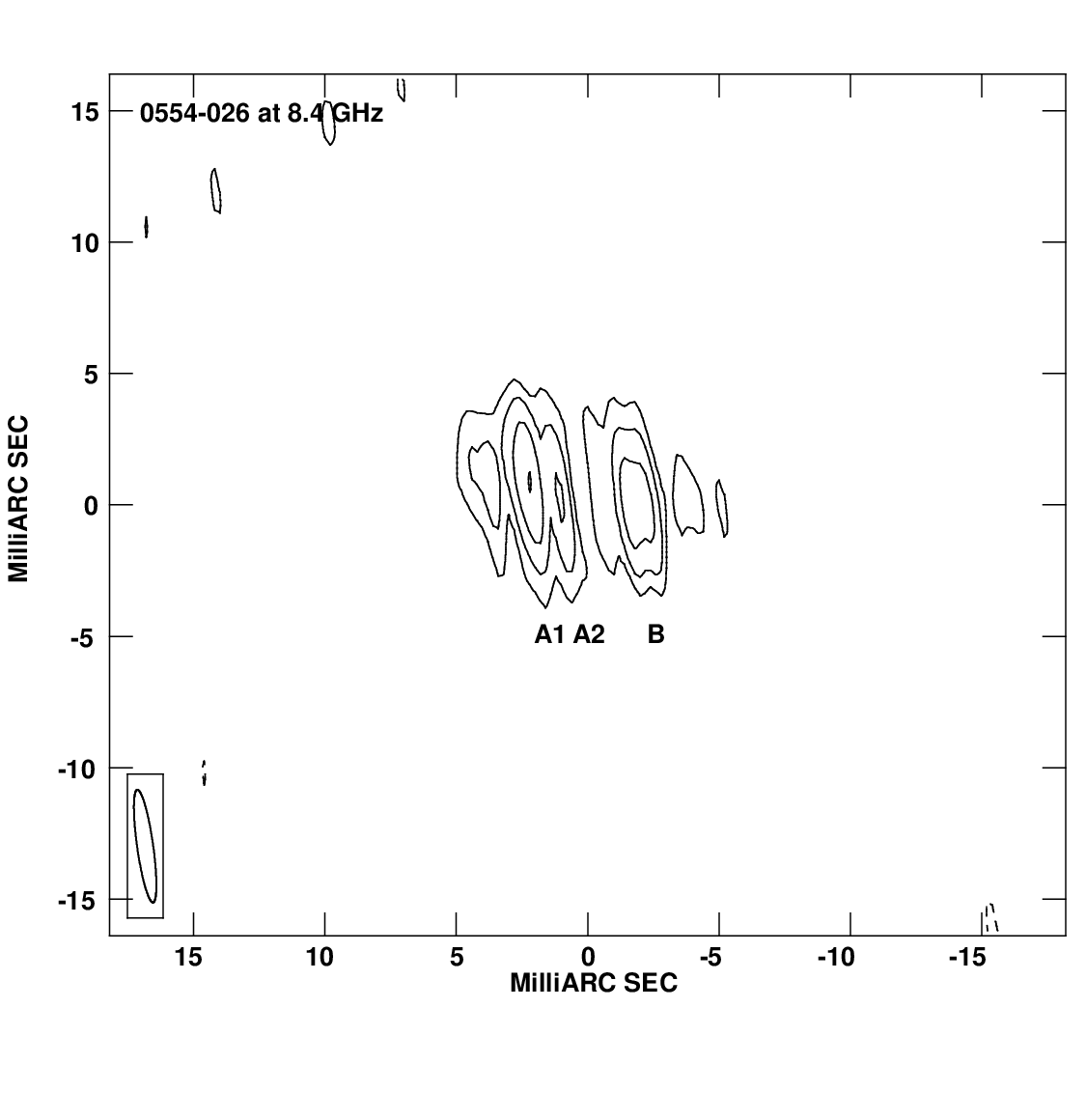}
      \caption{0554$-$026 at 8.4 GHz, the restoring beam is $4.3\times 0.6$ mas in PA $8^{\circ}$, the rms noise
      on the image is 0.6 mJy/beam, the lowest contour is 3$\sigma$, the peak flux density is 15 mJy/beam.}
      \label{0554X}
   \end{figure}

This is a galaxy with $m_{V}$=18.5 and redshift 0.235 (de Vries et
al. 2000). Both VLBI images (Fig. \ref{0554S} and Fig.
\ref{0554X}) presented here show a structure where a series of
well resolved components are aligned in the East-West direction.
There is no indication of a possible core candidate. Based on
these data a proper classification of this source is not possible,
although it may be still considered a CSO candidate.

\subsection{\object{0602+780} (\object{JVAS J0610+7801}, \object{NVSS J061024+78013})}

\begin{figure}
      \includegraphics[width=6cm]{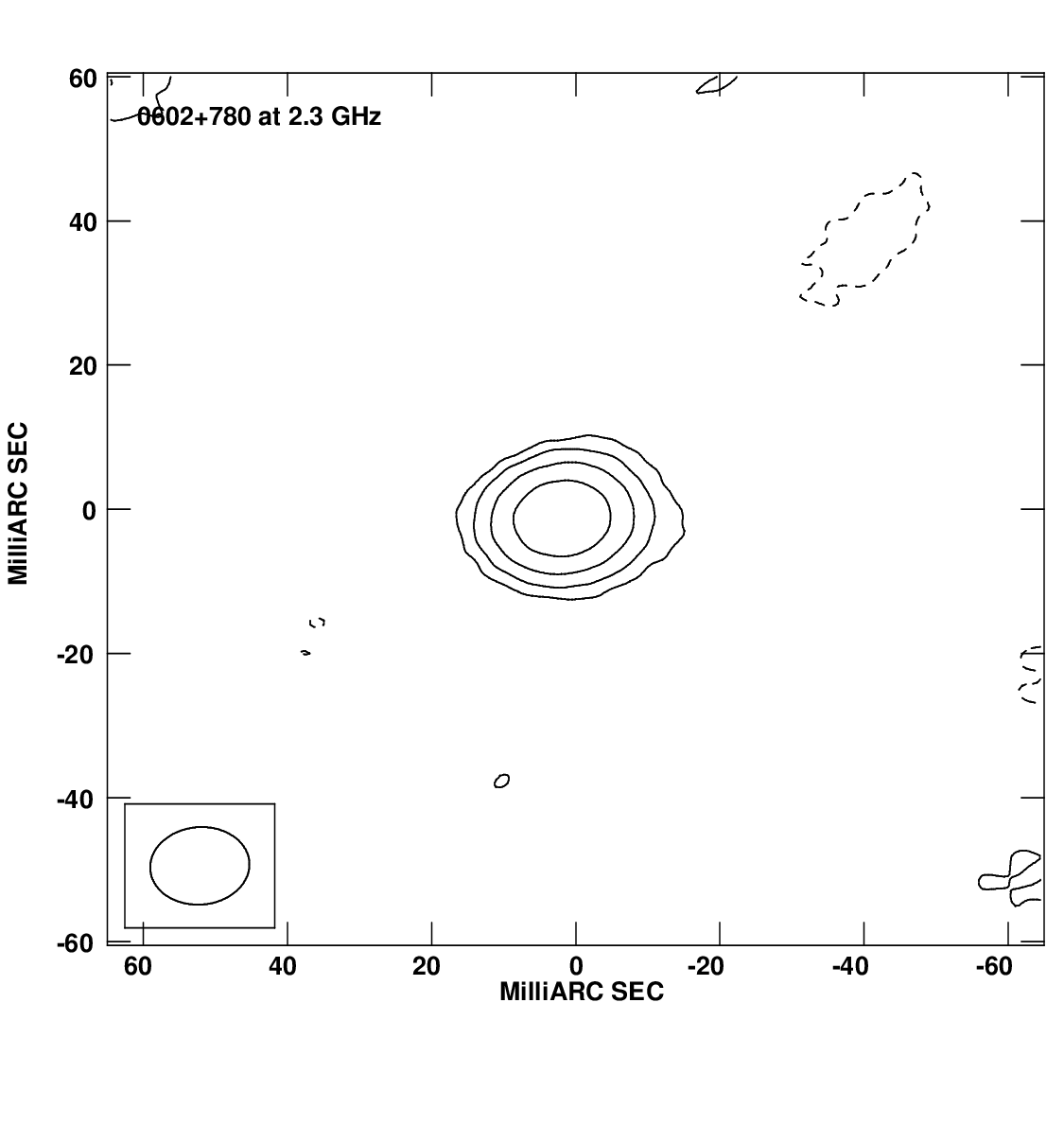}
      \caption{0602+780 at 2.3 GHz, the restoring beam is $13.8\times10.7$ mas in PA $-85.5^{\circ}$, the rms
      noise on the image is 1.2 mJy/beam, the lowest contour is 2.5$\sigma$, the peak flux density is 44 mJy/beam.}
      \label{0602S}
   \end{figure}

\begin{figure}
      \includegraphics[width=6cm]{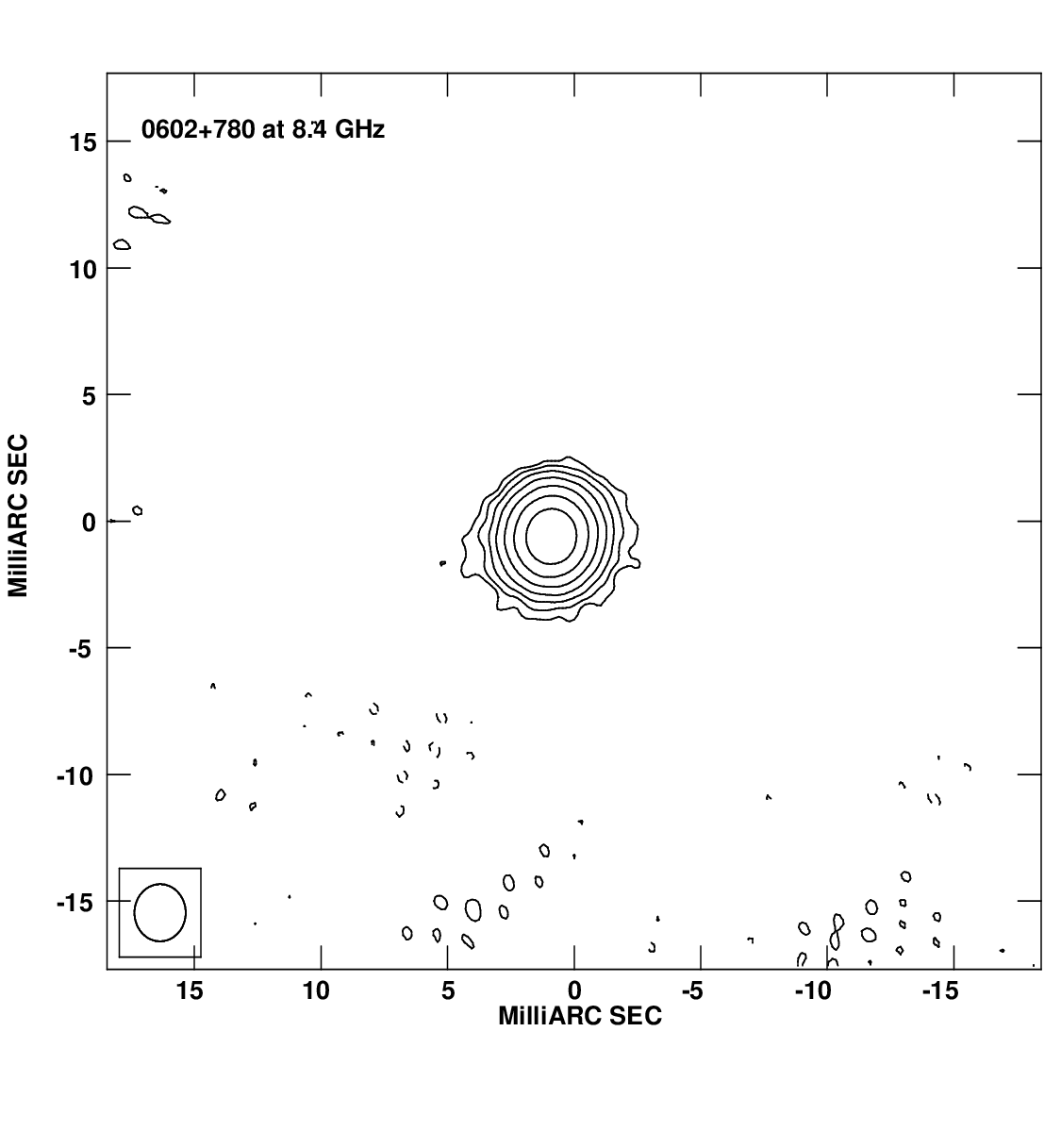}
      \caption{0602+780 at 8.4 GHz, the restoring beam is  $2.3\times2.0$ mas in PA $-0.6^{\circ}$, the rms noise
      on the image is 0.6 mJy/beam, the lowest contour is 1.3$\sigma$, the peak flux density is 94 mJy/beam.}
      \label{0602X}
   \end{figure}

This object is identified as a very faint galaxy by Stanghellini
et al. (1993). In the radio, the JVAS flux density at 8.4 GHz is
124.1 mJy and the source is a suitable MERLIN phase calibrator
(Wilkinson et al. 1998). In the VCS it is found to be point
source. Our images at both frequencies show a point-like source
(Fig. \ref{0602S} and Fig. \ref{0602X}), with a rather inverted
spectrum ($\alpha = -0.57$). About 15\% of the JVAS flux density
is not accounted for in the VLBI image of Fig.~\ref{0602X},
possibly related to a jet component completely resolved out by the
present observations at both frequencies.

\subsection{\object{PKS 0802+212} (\object{JVAS J0805+2106}, \object{NVSS J080538+210651})}

\begin{figure}
      \includegraphics[width=6cm]{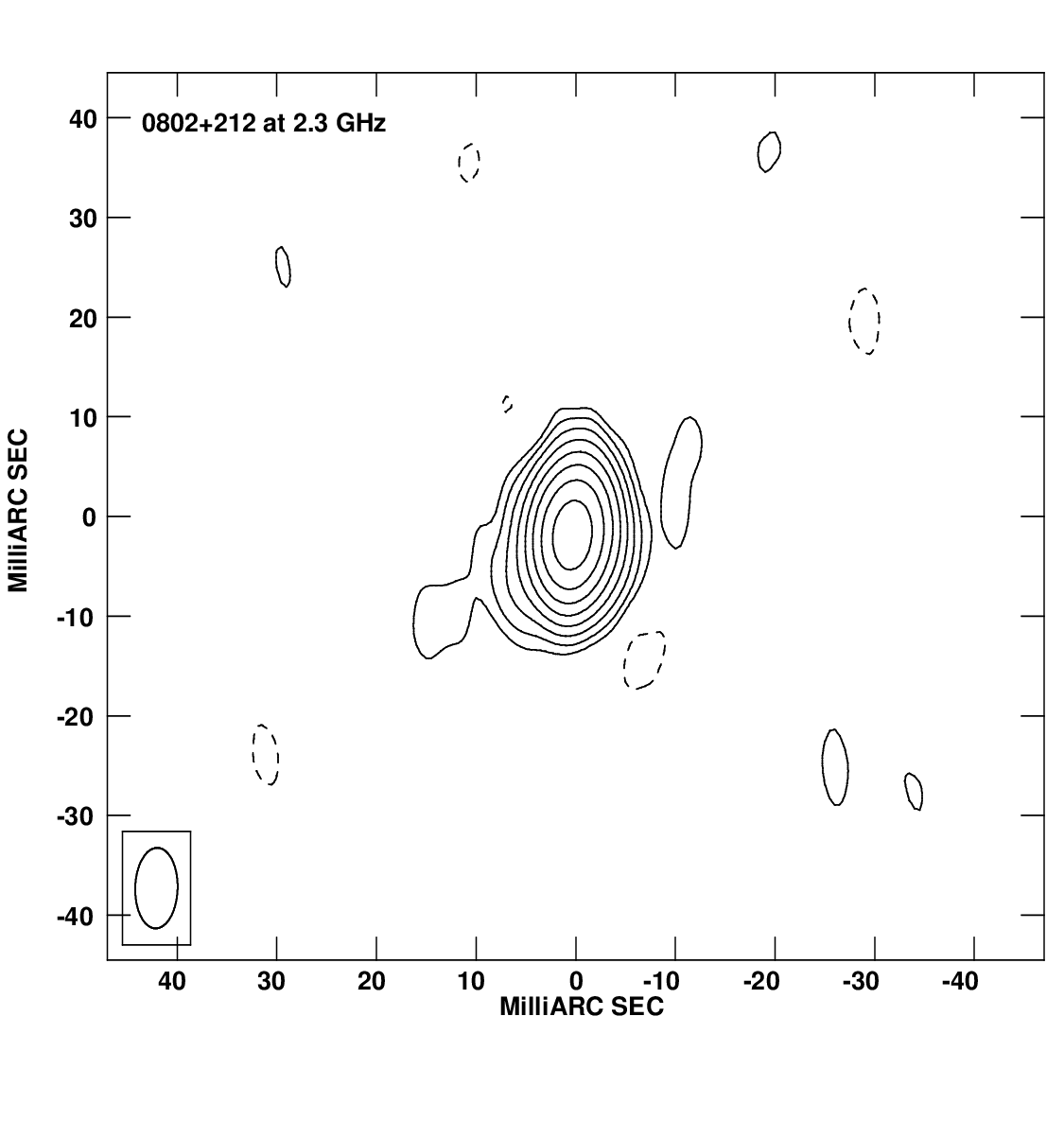}
      \caption{0802+212 at 2.3 GHz, the restoring beam is $8.1\times4.3$ mas in PA $-1.5^{\circ}$, the rms noise
      on the image is 0.8 mJy/beam, the lowest contour is 3.7$\sigma$, the peak flux density is 618 mJy/beam.}
      \label{0802S}
   \end{figure}

\begin{figure}
      \includegraphics[width=6cm]{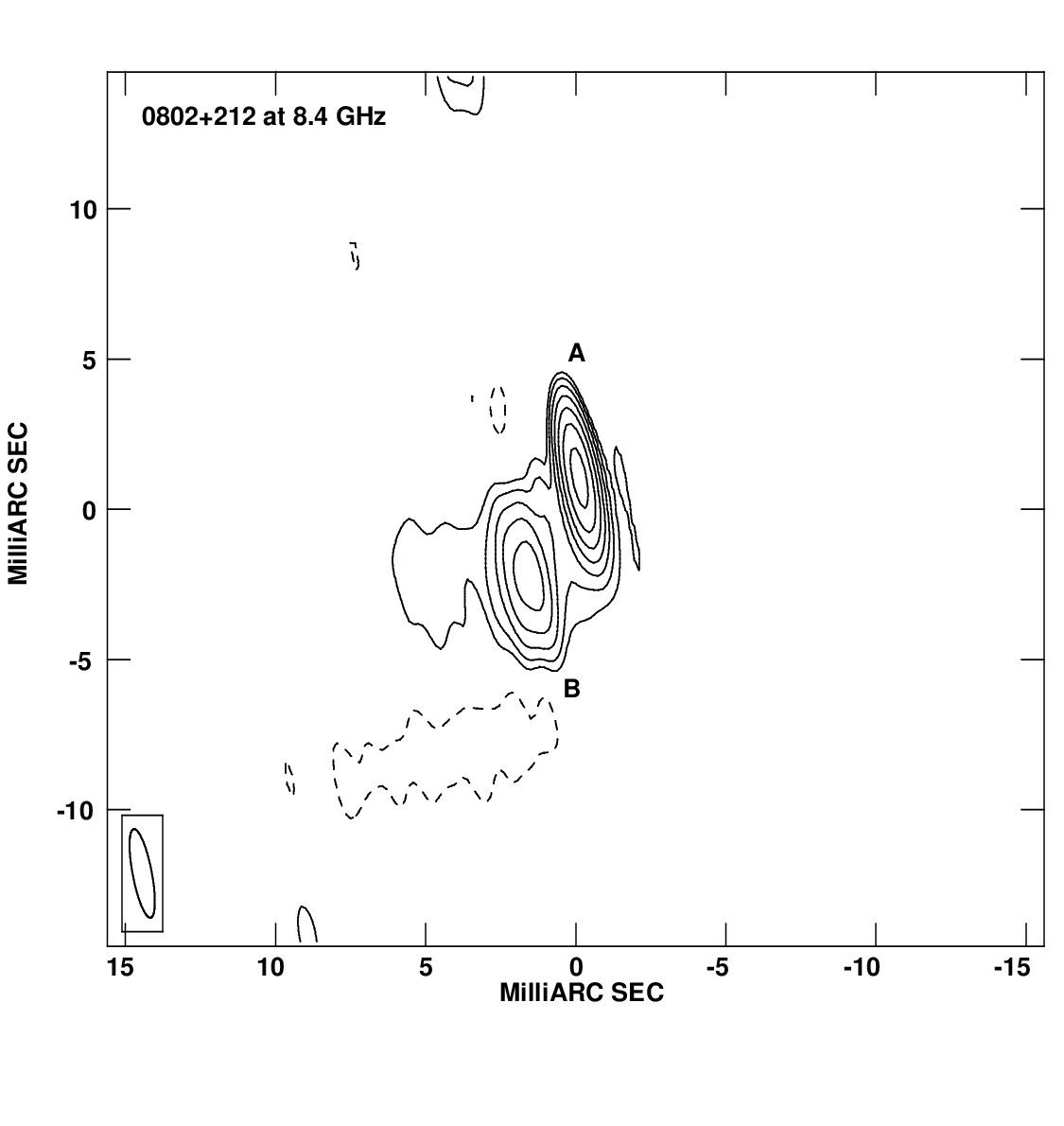}
      \caption{0802+212 at 8.4 GHz, the restoring beam is $3.1\times0.7$ mas in PA $10.7^{\circ}$, the rms noise
      on the image is 1.5 mJy/beam, the lowest contour is 2$\sigma$, the peak flux density is 264 mJy/beam.}
      \label{0802X}
   \end{figure}

The source is identified as a galaxy with $m_{R}$=22.5 (de Vries
et al. 2000). In the radio, the JVAS flux density at 8.4 GHz is
348.8 mJy and the source is a suitable MERLIN phase calibrator
(Wilkinson et al. 1998). The source was detected with a total flux
density of 0.22 Jy at 22 GHz by Ter\"asranta et al. (2001). Images
at 2.3 GHz and at 8.4 GHz are available from the VCS. Our VLBI
image at 2.3 GHz (Fig. \ref{0802S}) shows a marginally resolved
object, whose structure is well highlighted at 8.4 GHz (Fig.
\ref{0802X}) where a compact core-jet morphology is visible. Both
our images are consistent with the VCS results. Indeed the total
flux density accounted for in the 8.4 GHz image exceeds the VLA
measurement of the JVAS by about 17\%. There are also other
signatures of a similar flux density variability if we compare the
total flux density detected at 1.4 GHz in the NVSS and the FIRST
(about 18\% weaker). All these pieces of information suggest a
core-jet classification for this source.

\subsection{\object{PKS 0904+039} (\object{NVSS J090641+034242})}

The source is identified with a galaxy, $m_{I}$=22.1, which
appears slightly extended in an I band image. This object seems to
be located in a small group of objects (presumably galaxies, de
Vries et al. 2000).

In the radio, some extended emission to the South is visible in
the NVSS image. However, the FIRST image of the same area shows
that it is likely due to an unrelated background/foreground
object. Indeed PKS 0904+039 is slightly resolved in the FIRST
survey and with a total flux density exceeding by about 10\% the
one measured in the NVSS.

\begin{figure}
      \includegraphics[width=6cm]{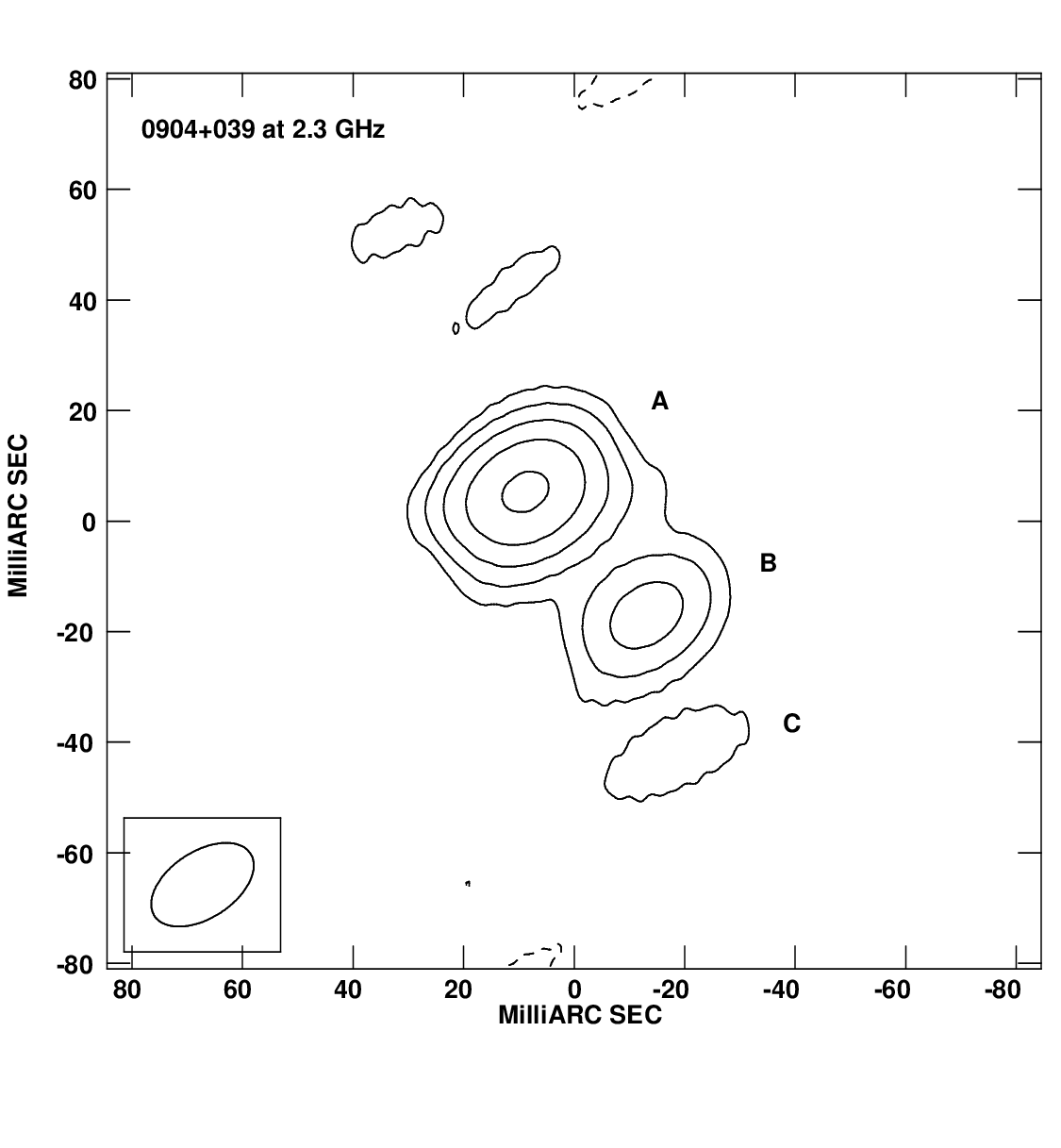}
      \caption{0904+039 at 2.3 GHz, the restoring beam is $20.7\times12$ mas in PA $-57^{\circ}$, the rms noise
      on the image is 5.9 mJy/beam, the lowest contour is 2.5$\sigma$, the peak flux density is 273 mJy/beam.}
      \label{0904S}
   \end{figure}

\begin{figure}
      \includegraphics[width=6cm]{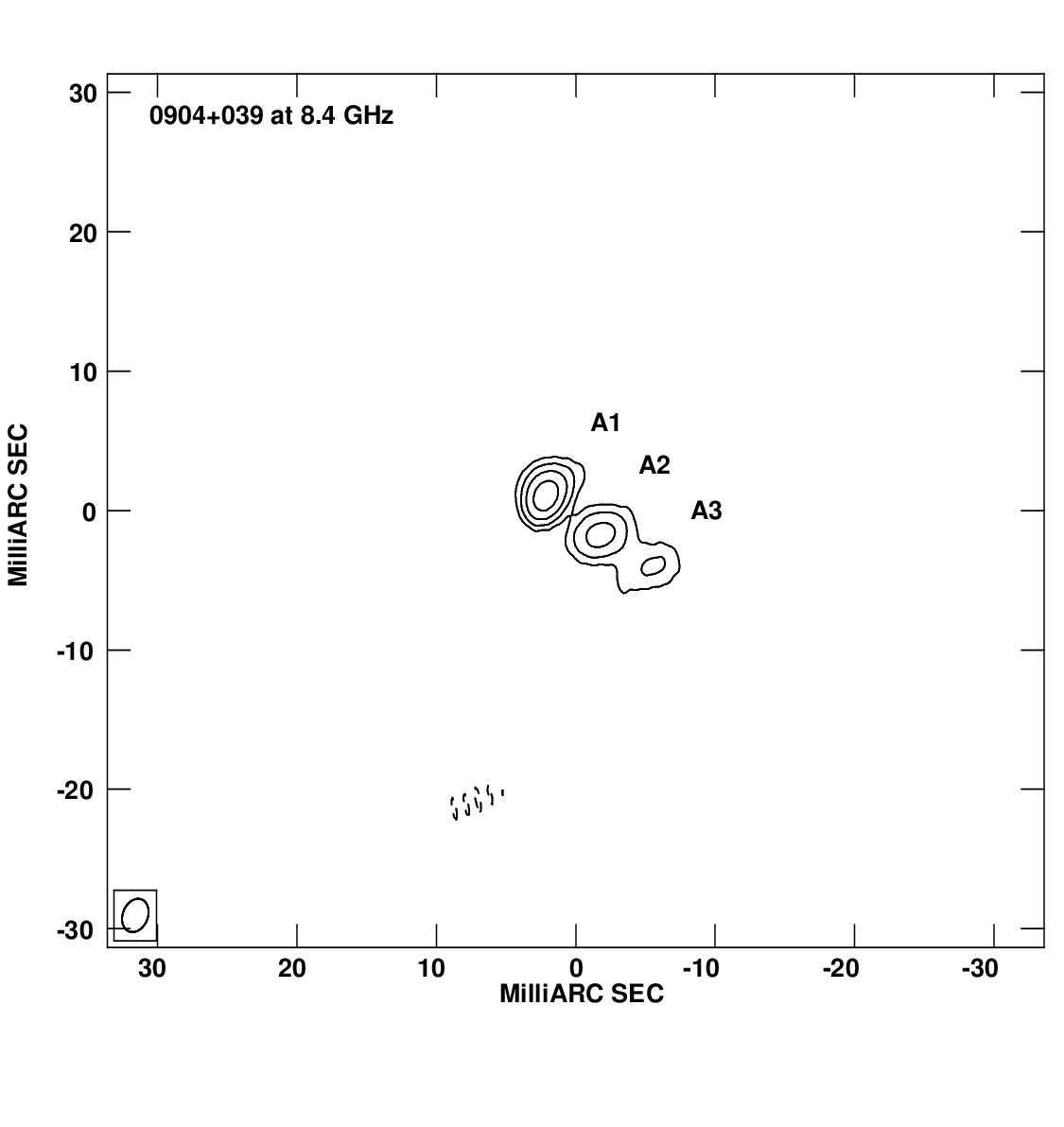}
      \caption{0904+039 at 8.4 GHz, the restoring beam is $2.5\times1.8$ mas in PA $-21.3^{\circ}$, the rms noise
      on the image is 0.4 mJy/beam, the lowest contour is 3.5$\sigma$, the peak flux density is 17 mJy/beam.}
      \label{0904X}
   \end{figure}

The VLBI image at 2.3 GHz (Fig. \ref{0904S}) shows a series of
resolved components roughly aligned p.a. $-140^\circ$ with a
possible additional weak component to the South. Our image at 8.4
GHz (Fig.~\ref{0904X}) detects three blobs in which component A in
Fig.~\ref{0904S} is resolved. None of them is particularly compact
enough to be considered as the source core. More than 50\% of the
total flux density at this frequency (cf. Stanghellini et al.
1998) is not accounted for in the VLBI image. Conversely, at 2.3
GHz the total flux density accounted for in our VLBI image
slightly exceeds the literature data. We can consider this source
a CSO candidate.

\subsection{\object{PKS 0914+114} (\object{NVSS J091716+111336})}

\begin{figure}
      \includegraphics[width=6cm]{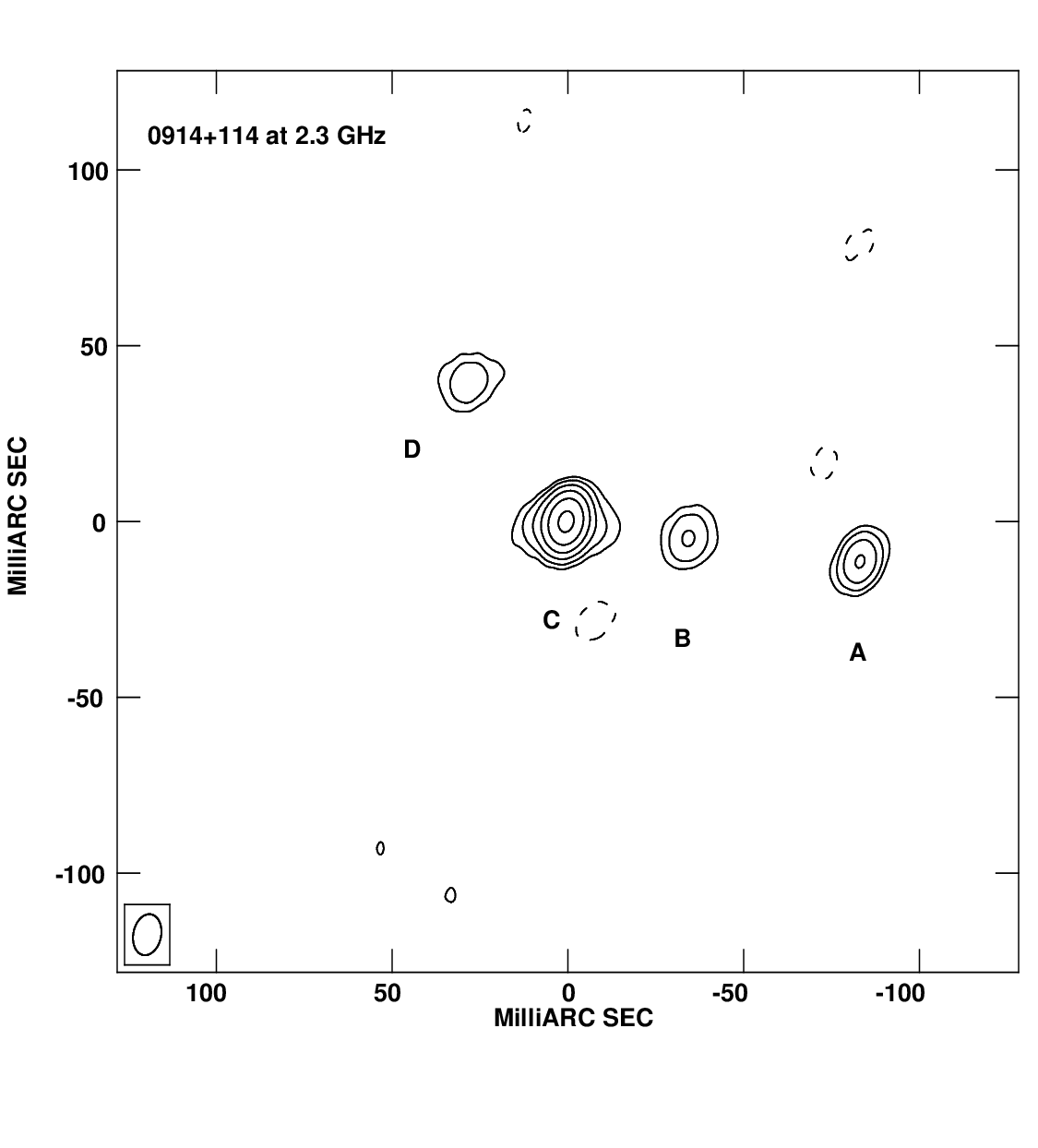}
      \caption{0914+114 at 2.3 GHz, the restoring beam is  $12\times7.8$ mas in PA $-11.2^{\circ}$, the rms noise
      on the image is 1.4 mJy/beam, the lowest contour is 3.6$\sigma$, the peak flux density is 126 mJy/beam.}
      \label{0914S}
   \end{figure}

\begin{figure}
      \includegraphics[width=6cm]{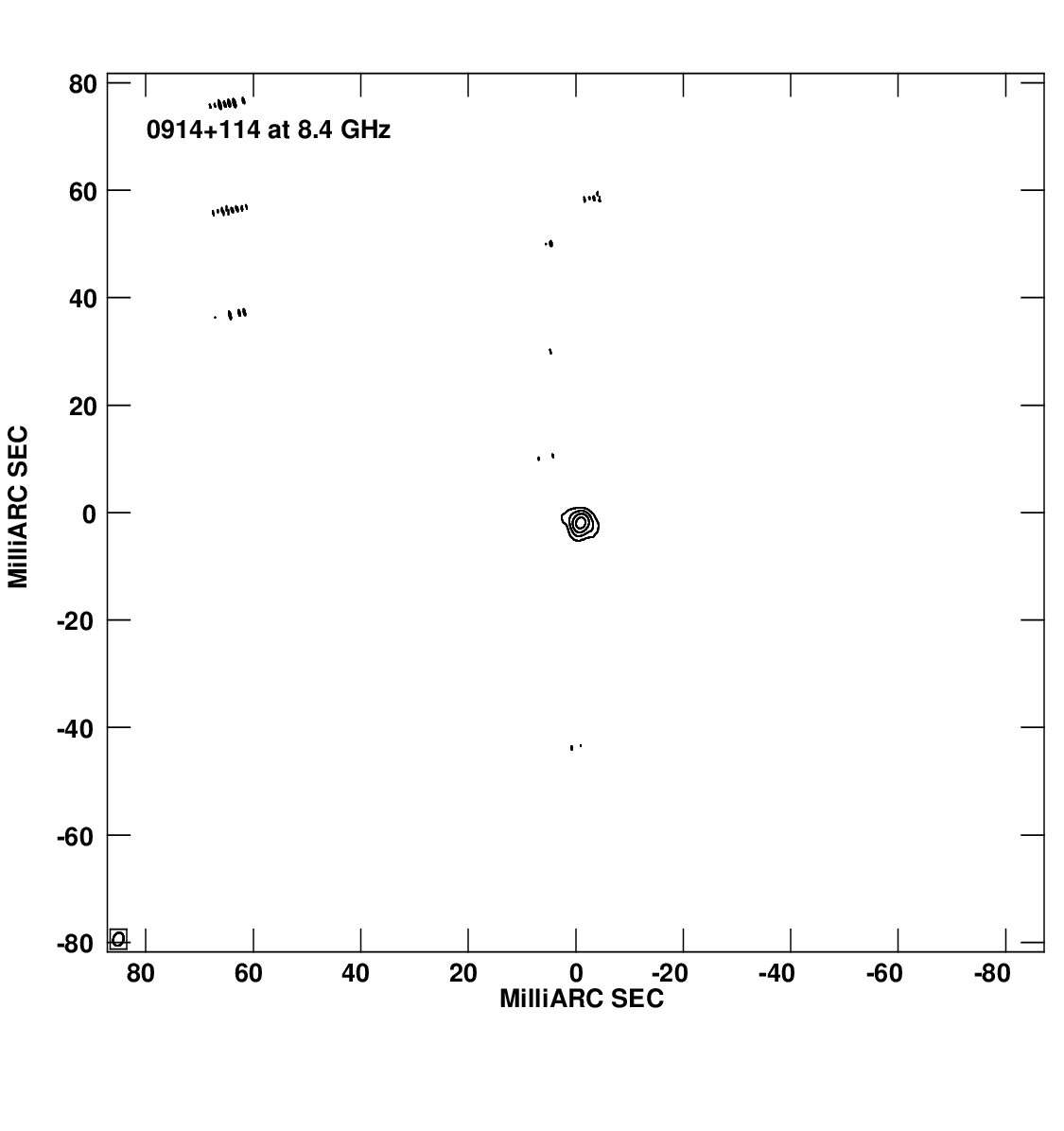}
      \caption{0914+114 at 8.4 GHz, the restoring beam is $2.5\times1.9$ mas in PA $-18^{\circ}$, the rms noise
      on the image is 0.3 mJy/beam, the lowest contour is 2.7$\sigma$, the peak flux density is 20 mJy/beam.}
      \label{0914X}
   \end{figure}

This source is associated with a galaxy, $m_{r}$=20.0
(Stanghellini et al. 1993), z=0.178 (de Vries et al. 1998). Our
VLBI image at 2.3 GHz (Fig.~\ref{0914S}), accounting for the whole
source flux density at this frequency, shows four well separated
components, with no obvious interpretation of the structure. Only
one of them, likely component C, is detected at 8.4 GHz
(Fig.~\ref{0914X}). About 50\% of the total flux density at this
frequency (cf. Stanghellini et al. 1998) is missing in the image.
If it is the component C, its spectral index would be $\alpha =
1.73$, slightly steeper than the optically thin spectral index
($\alpha = 1.6$) derived from VLA data. Given this steep spectral
index and the source structure seen in our data, we can consider
this object a CSO candidate.

\subsection{\object{B3 1133+432} (\object{NVSS J113555+425844})}

\begin{figure}
      \includegraphics[width=6cm]{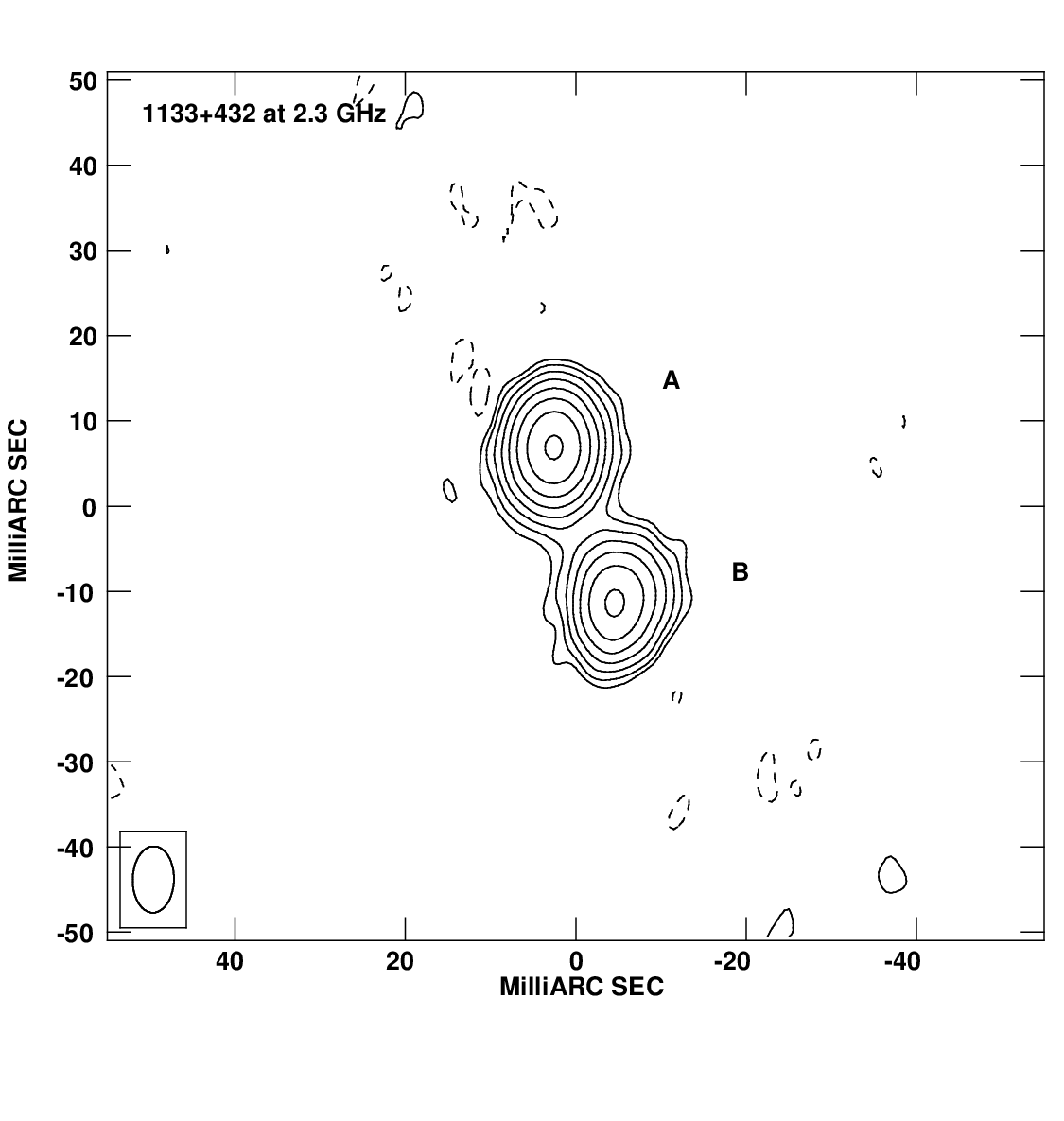}
      \caption{1133+432 at 2.3 GHz, the restoring beam is $7.8\times4.8$ mas in PA $-1.3^{\circ}$, the rms noise
      on the image is 1.7 mJy/beam, the lowest contour is 2.4$\sigma$, the peak flux density is 560 mJy/beam.}
      \label{1133S}
   \end{figure}

\begin{figure}
      \includegraphics[width=6cm]{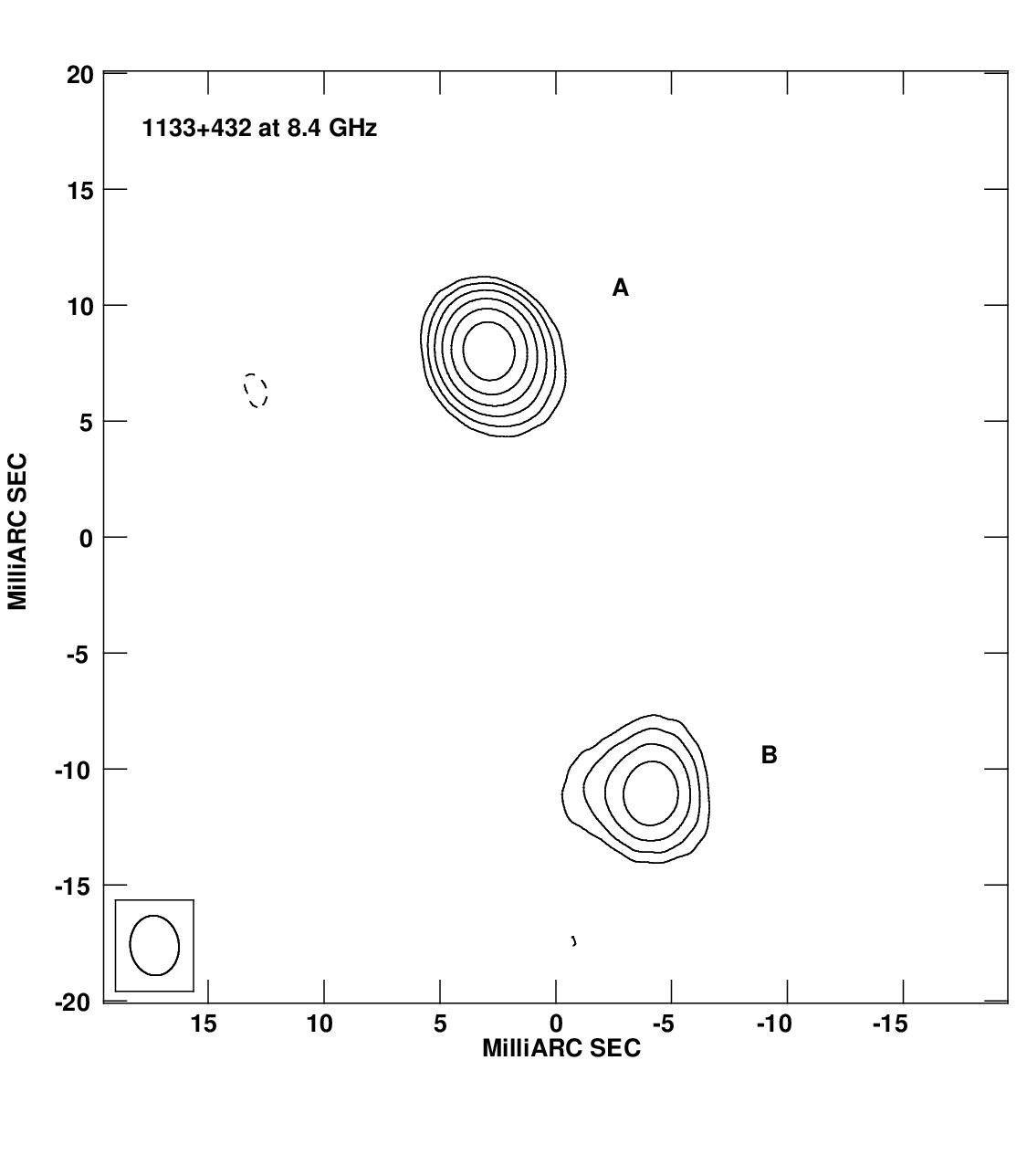}
      \caption{1133+432 at 8.4 GHz, the restoring beam is $2.6\times2.1$ mas in PA $6.6^{\circ}$, the rms noise
      on the image is 0.6 mJy/beam, the lowest contour is 3.3$\sigma$, the peak flux density is 117 mJy/beam.}
      \label{1133X}
   \end{figure}

This is an empty field in the optical (Stickel et al. 1994). Our
VLBI images at 2.3 GHz and at 8.4 GHz (Fig.~\ref{1133S} and
Fig.~\ref{1133X}) show a double source, with both components
resolved by the present observations. This source has been studied
in detail by Dallacasa et al. (2002) at 1.7 GHz and by Orienti et
al. (2004) at 5.0 GHz and at 8.4 GHz with the VLBA, as part of a
study on the Compact Steep Spectrum sources from the B3-VLA sample
selected by Fanti et al. (2001). Our results are consistent with
the aforementioned papers, although our 8.4 GHz image could not
account for about 35\% of the total flux density. The spectral
indices we derived for our components between 2.3 and 8.4 GHz
($\alpha = 1.23$ and $\alpha = 1.65$) are steeper than those
derived by Orienti et al. (2004) between 5.0 and 8.4 GHz due to
the smaller fraction of flux density accounted for in our 8.4 GHz
image.

Despite the lack of a core detection, which would be one
requirement for a proper CSO classification, we consider this
object as a good CSO source, since we interpret the overall
morphology as a very compact, lobe dominated double; to detect the
core we would need a much higher dynamic range and possibly a
better UV-coverage at 8.4 GHz.

\subsection{\object{1824+271} (\object{J1826+2707})}

\begin{figure}
      \includegraphics[width=6cm]{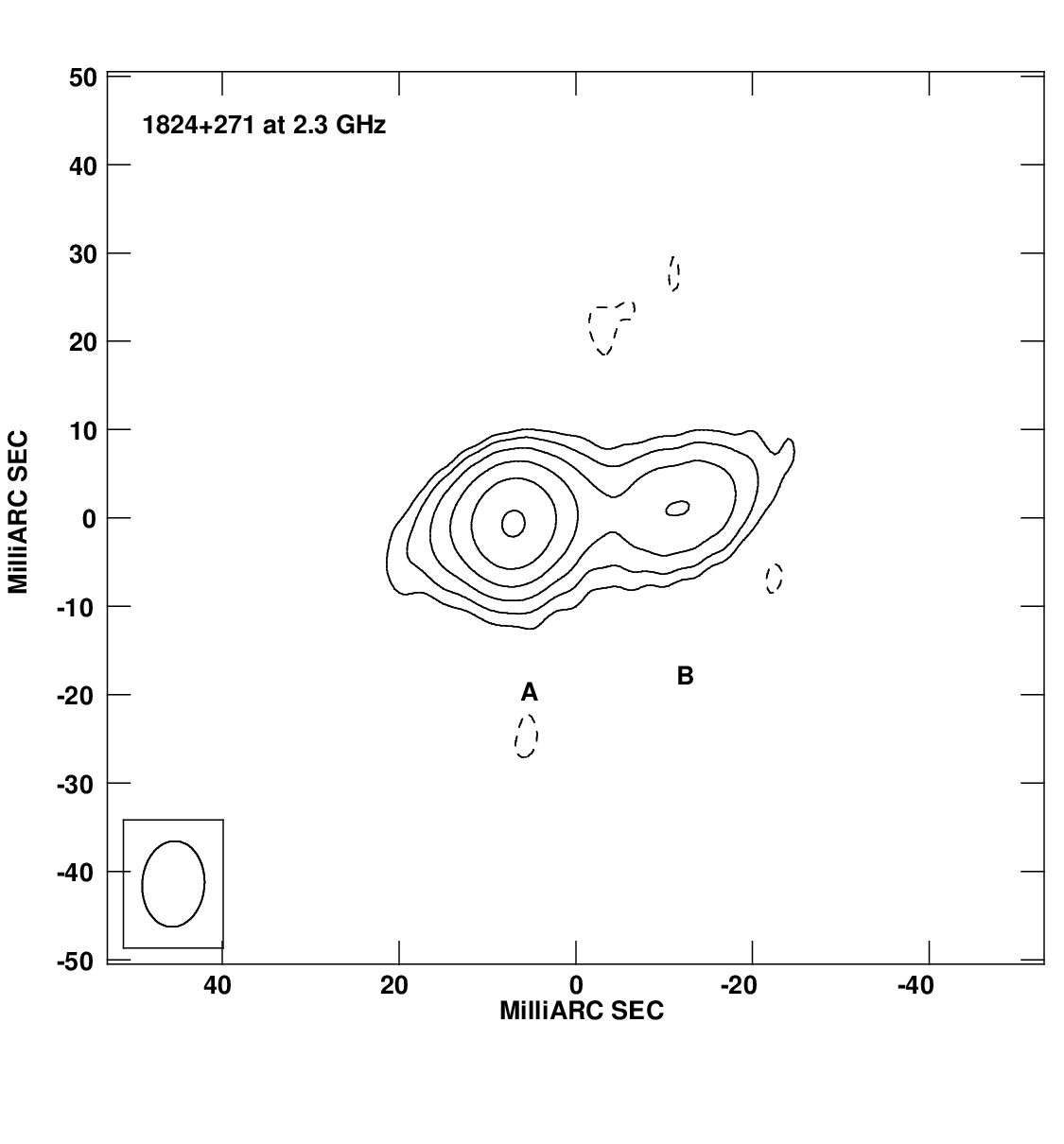}
      \caption{1824+271 at 2.3 GHz, the restoring beam is $9.5\times6.7$ mas in PA $-3.6^{\circ}$, the rms noise
      on the image is 1.1 mJy/beam, the lowest contour is 3.6$\sigma$, the peak flux density is 137 mJy/beam.}
      \label{1824S}
   \end{figure}

\begin{figure}
      \includegraphics[width=6cm]{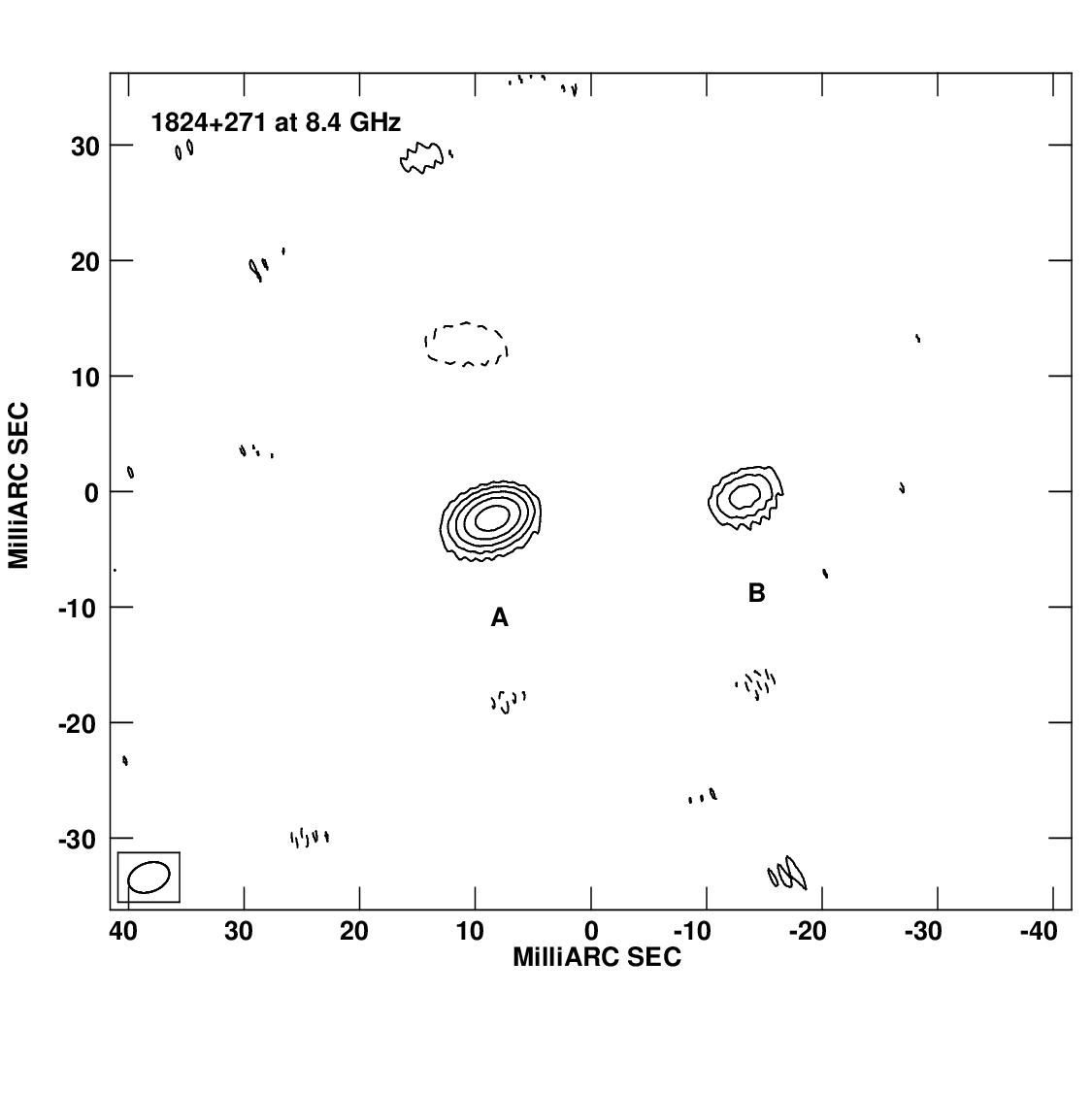}
      \caption{1824+271 at 8.4 GHz, the restoring beam is $2.7\times1.9$ mas in PA $-70^{\circ}$, the rms noise
      on the image is 0.4 mJy/beam, the lowest contour is 2.5$\sigma$, the peak flux density is 24 mJy/beam.}
      \label{1824X}
   \end{figure}

It is associated with a faint object with $m_{R}$=22.9 (O'Dea et
al. 1990). Given such weak magnitude its optical counterpart does
not have a morphological classification (G/Q) although galaxies
are more frequently found than quasars associated with GPS
sources. In the radio, our 2.3 and 8.4 GHz VLBI images
(Fig.~\ref{1824S} and Fig.~\ref{1824X}) show a double structure
roughly aligned in the East-West direction; both components have
steep spectra ($\alpha = 1.37$ for the brighter A component,
$\alpha = 1.63$ for the weaker B component) and are consistent
with the overall steep radio spectrum ($\alpha = 1.4$) derived
from the literature data. In fact a flux density of 50 mJy at
8.085 GHz and of 230 mJy at 2.695 GHz are found in the NED
database. These flux densities indicate that while the whole
source flux density is accounted for in our 2.3 GHz VLBI image,
some fraction of the flux density may be missing in our 8.4 GHz
VLBI image.

The source structure is consistent with the expectation of a CSO
seen in low dynamic range images, and therefore we suggest this
classification.

\subsection{\object{PKS 2121$-$014} (\object{NVSS J212339$-$011234})}

This is a galaxy with $m_{R}$=23.3, which shows double structure
presumably indicating a case of close interaction (de Vries et al.
2000). The redshift for this system is 1.158 (Snellen et al.
1996).

\begin{figure}
      \includegraphics[width=6cm]{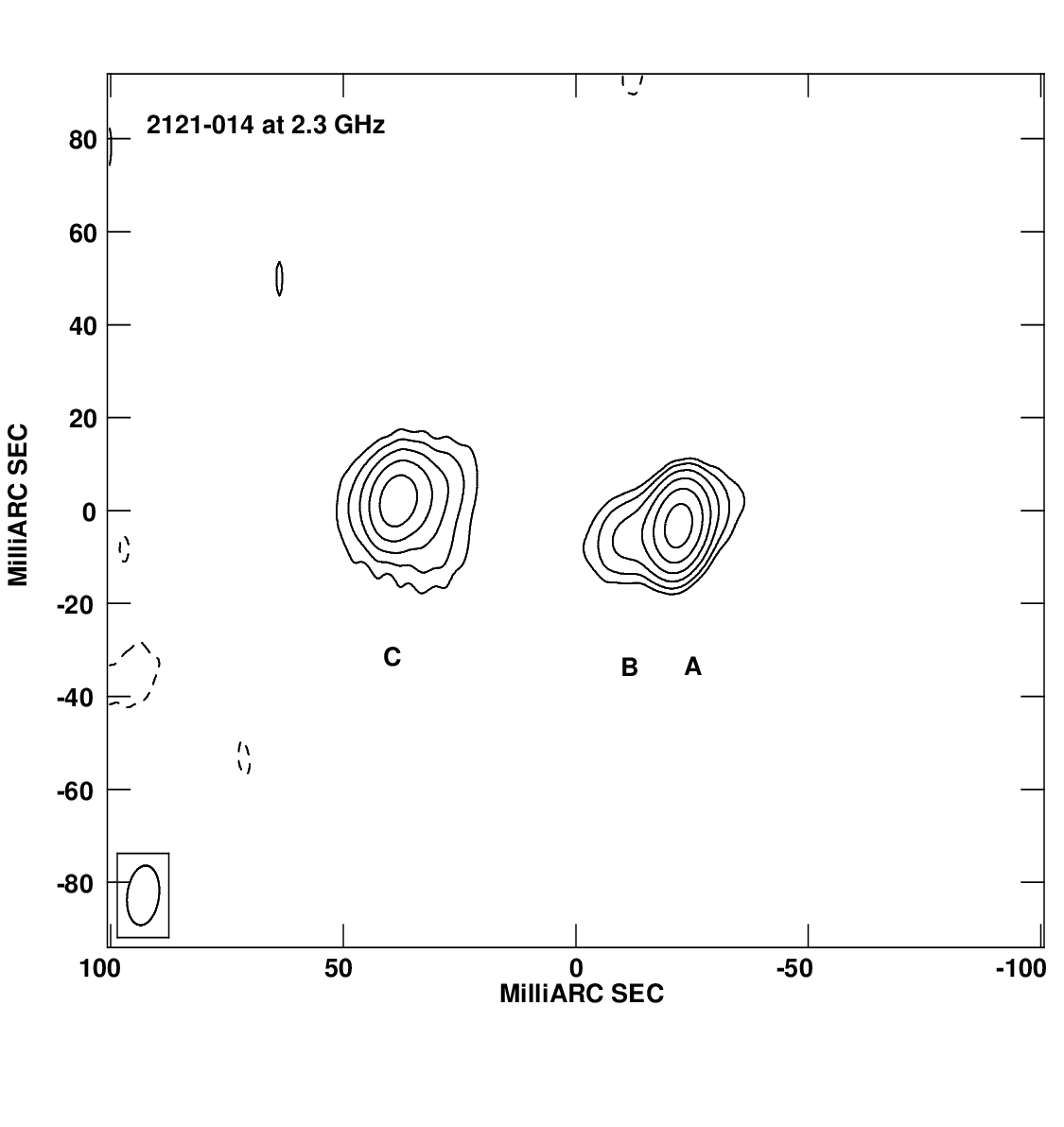}
      \caption{2121$-$014 at 2.3 GHz, the restoring beam is $12.9\times6.8$ mas in PA $-6.8^{\circ}$, the rms noise
      on the image is 2.3 mJy/beam, the lowest contour is 2.6$\sigma$, the peak flux density is 276 mJy/beam.}
      \label{2121S}
   \end{figure}

\begin{figure}
      \includegraphics[width=6cm]{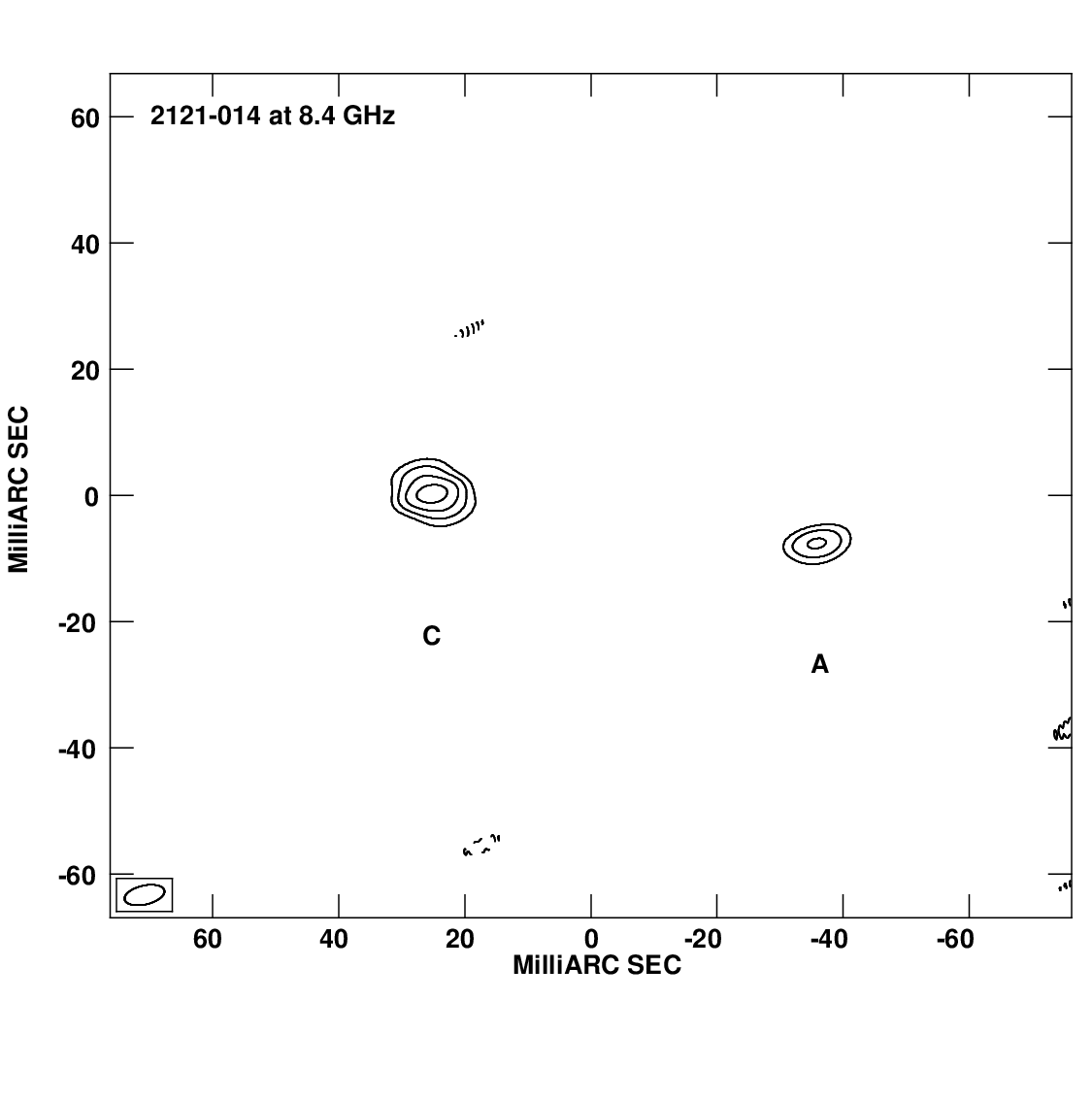}
      \caption{2121$-$014 at 8.4 GHz, the restoring beam is $6.4\times3.0$ mas in PA $-77^{\circ}$, the rms noise
      on the image is 1.5 mJy/beam, the lowest contour is 4$\sigma$, the peak flux density is 70 mJy/beam.}
      \label{2121X}
   \end{figure}

Our VLBI images exhibit a double structure with both components
well resolved. At 2.3 GHz (Fig.~\ref{2121S}) a second component
may be added to the main one (component A), possibly representing
a sort of tail/lobe emission. However, the component A is quite
weak at 8.4 GHz (Fig.~\ref{2121X}) with a spectral index of 1.43,
indicating it is not the core of the source. Surprisingly,
component C has a mid-steep spectrum ($\alpha =0.65$), despite its
considerably larger size at 2.3 GHz. It is also possible that in
the early data reduction the symmetrization caused by the fringe
fitting process has been wrongly resolved, resulting in a swap of
the position of the two components.

As in the case of 1824+271, we consider this source as a CSO.

\subsection{\object{PKS 2322$-$040} (\object{NVSS J232510$-$034446})}

This is a galaxy with $m_{R}$=23.5, which appears to be a regular
elliptical and possibly a member of a small group of galaxies (de
Vries et al. 2000).

\begin{figure}
      \includegraphics[width=6cm]{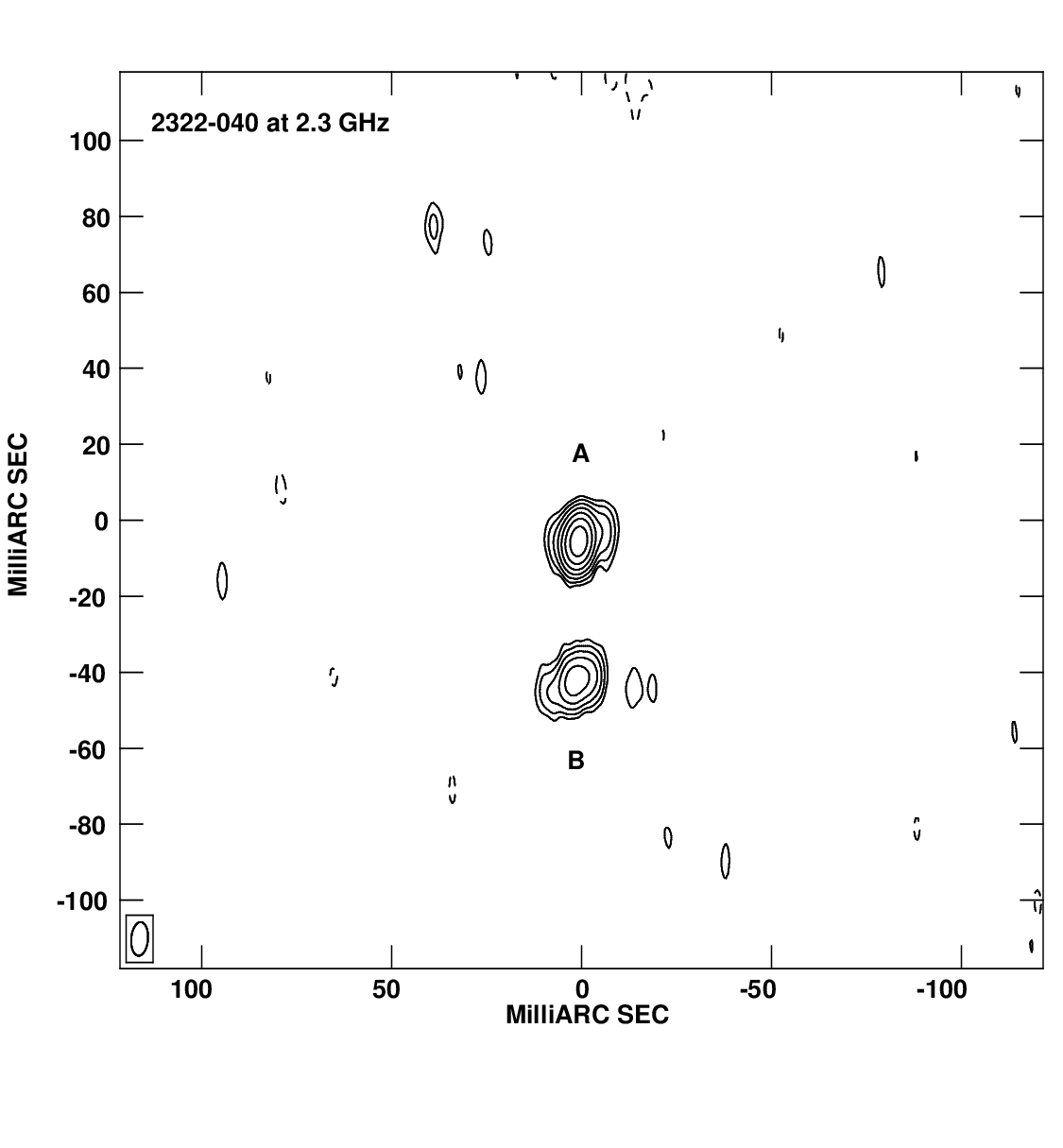}
      \caption{2322$-$040 at 2.3 GHz, the restoring beam is $8.8\times4.4$ mas in PA $-4.5^{\circ}$, the rms noise
      on the image is 1.6 mJy/beam, the lowest contour is 2.5$\sigma$, the peak flux density is 541 mJy/beam.}
      \label{2322S}
   \end{figure}

\begin{figure}
      \includegraphics[width=6cm]{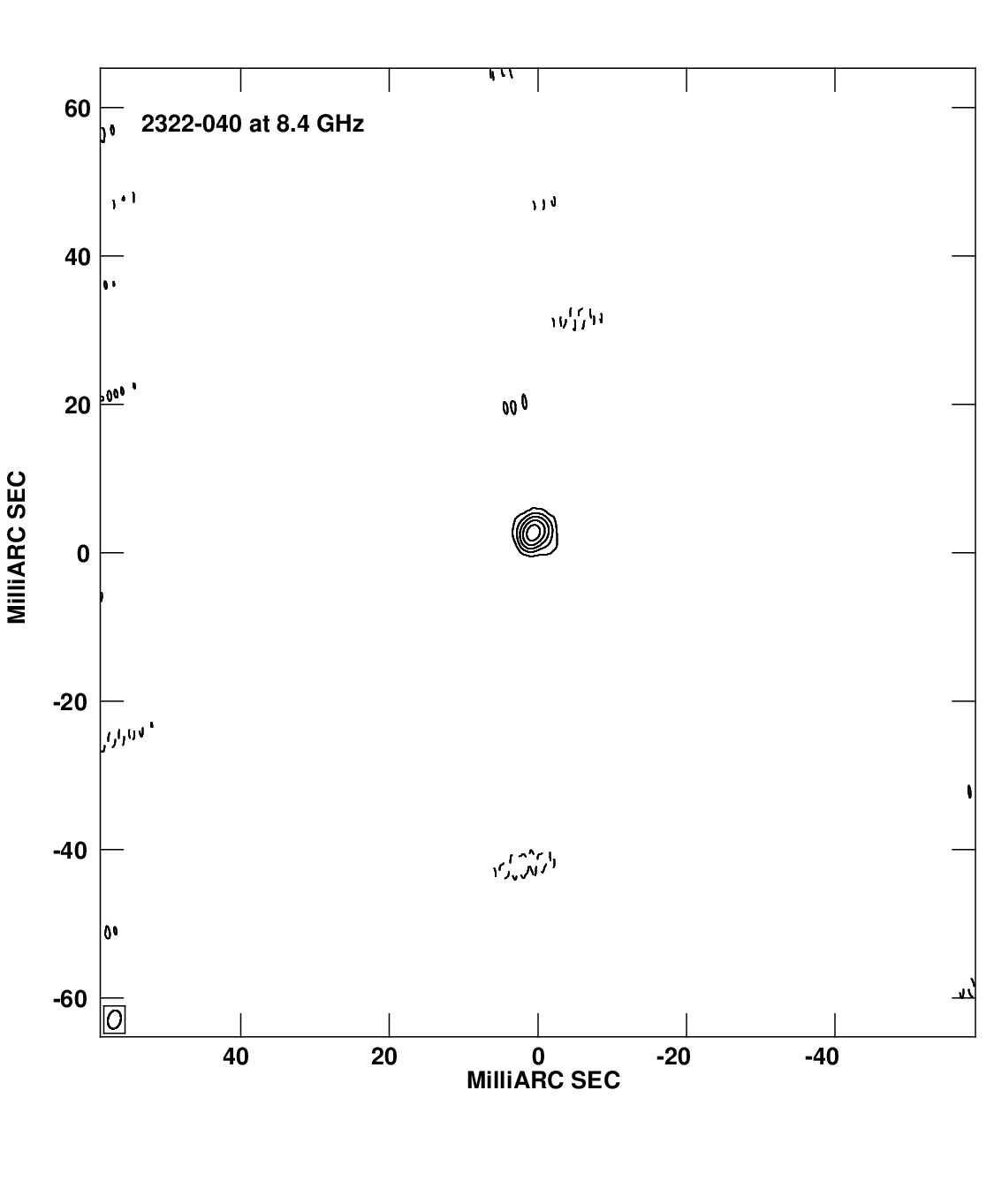}
      \caption{2322$-$040 at 8.4 GHz, the restoring beam is $2.5\times1.7$ mas in PA $-13^{\circ}$, the rms noise
      on the image is 1.6 mJy/beam, the lowest contour is 1.9$\sigma$, the peak flux density is 125 mJy/beam.}
      \label{2322X}
   \end{figure}

Our VLBI images show a double structure at 2.3 GHz
(Fig.~\ref{2322S}), while only a single component is detected at
8.4 GHz. This component is slightly resolved (Fig.~\ref{2322X}).
The whole flux density is accounted for in the 2.3 GHz image,
while at 8.4 GHz we can account for 166 mJy only, while a flux
density of 240 mJy is reported at this frequency in the NED
database. The missing flux density may be in the secondary
component 'B' which is lost in our 8.4 GHz image due to poor UV
coverage in the North-South direction (there is a negative beam
sidelobe at the position of component 'B' in Fig.22 preventing us
from collecting enough flux density in that area).

The spectral index for the component A is 1.1 (consistent with the
spectral index of the whole source $\alpha=1.16$, as measured from
the NED values), indicating it is more likely a lobe rather than
the core of the source.

The source 2322$-$040 can be considered a CSO candidate.

\section{Discussion}

CSOs are defined as very compact radio sources (a few hundreds of
pc in size) whose radio emission is dominated by a pair of
hot-spots and lobes lying on the opposite side of a core. The
overall source spectrum has a convex shape as commonly found in
GPS radio sources, and most if not all CSOs are found among GPSs
or at least among peaked spectrum radio sources.

Ideally, a core component, with a flatter (often also inverted)
spectrum than the hot-spots and lobes must be identified before a
compact radio source can be confirmed as a CSO. For some CSOs with
a jet axis very close to the plane of sky the core may be Doppler
dimmed and become so weak as to be undetectable. However the CSO
classification may be still valid if there are symmetric
edge-brightened hot spots and/or extended lobes (Taylor \& Peck
2003). Examples can be found in J1734+0926 (Peck \& Taylor 2000)
and in 1133+432, 1824+271 and 2121$-$014 presented in this paper.

The core of CSOs in some cases can become brighter at high
frequencies, e.g. CSO 1946+708 (Taylor \& Vermeulen 1997) due to a
turnover frequency substantially higher than the lobes and the
hot-spots. It is then possible to detect the core in CSOs at
frequencies higher than 8.4 GHz. On the other hand, a few CSOs
possess a relatively bright core at centimeter wavelengths (Peck
\& Taylor 2000), and they could be affected by Doppler boosting to
some extent.

The flux densities at 2.3 GHz from our VLBI images account for the
whole radio source emission as extrapolated by the 2.7 GHz
literature data. All individual deviations can be referred to the
amplitude calibration error. Most of the sources have steep radio
spectra above 2.3 GHz, and the component spectral indices we found
may be steeper than those reported by de Vries et al. (1997) since
some amount of flux density is not accounted for in our VLBI
images at 8.4 GHz, possibly due to the lack of short UV spacings.

Concerning the sources we classify as CSO or candidate, none of
them has the core detected, possibly due to a limited image
dynamic range and/or to an intrinsically weak core. The spectral
index of the components detected in our image is generally very
steep, often  with $\alpha > 1$. This evidence coupled with the
fact that all the components are well resolved in our VLBI images
led us to consider them as possible CSO radio sources.

\section{Summary and conclusions}

Below we summarize the results and possible classification of
the sources.

1) All sources we observed have been detected at 2.3/8.4 GHz. For some
sources these are the first VLBI images allowing the study of the
morphology with parsec scale resolution.

2) The sources 1133+432, 1824+271 and 2121$-$014 have radio
emission coming from two well-resolved, steep spectrum regions
with comparable flux density and therefore can be classified as
bona-fide CSOs.

3) Other sources have a more complex structure: 0914+114 consists
of 4 components at 2.3 GHz but only one is detected at 8.4 GHz,
similarly 2322$-$040 is double at 2.3 GHz, but has a single
component detected at 8.4 GHz. The sources 0144+209, 0554$-$026
and 0904+039 are resolved into a series of small components at
both frequencies, all with rather steep spectra. Given the source
complexity our observations could not provide a UV-coverage good
enough to allow imaging of more details. We still consider these
five sources as CSO candidates.

4) Of the remaining three sources, two  (0159+839 and 0602+780)
have a single, inverted spectrum component detected, while the
third (0802+212) is marginally resolved at 2.3 GHz and shows a
typical core-jet structure in our 8.4 GHz image.

A sound statistical study will be possible once the sources from
the "bright" GPS samples have been well explored with VLBI.

Multi-frequency VLBI images are suited to do the spectral analysis
of components in more detail. Further work is required on the CSO
and CSO candidates presented here, and images at 5 GHz with a
better sensitivity are necessary to confirm their classification.

Our images of 1133+432, 1824+271 and 2121$-$014 can be used in
future studies aiming to measure hot-spot proper motion.

\begin{acknowledgements}

This work was supported by the Natural Science Foundation of China
(NSFC) under grant No.19973014, No.10373019 and the fund with No.
G1999075403. We thank the anonymous referee for important
comments, and R. Str\"om and Z.-Q Shen for helpful comments. L.X.
thanks Carlo Stanghellini for help with the proposal of this
project, and L. Gurvits, M. A. Garrett and B. Campbell for help
when L.X. was doing data calibration in JIVE. This research has
made use of the NASA/IPAC Extragalactic Database (NED) which is
operated by the Jet Propulsion Laboratory, Caltech, under contract
with NASA.

\end{acknowledgements}

\listofobjects

\clearpage

\clearpage

\clearpage

\clearpage

\clearpage

\end{document}